\documentclass[journal,onecolumn,12pt]{IEEEtran}
\usepackage[doublespacing,nodisplayskipstretch]{setspace}
\usepackage[utf8]{inputenc} 
\usepackage[T1]{fontenc}
\usepackage[colorlinks,pdfstartview=FitH,citecolor=blue,linkcolor=blue,urlcolor=blue]{hyperref}
\usepackage{url,comment}
\usepackage{ifthen}
\usepackage{graphicx}
\usepackage{amsmath,amssymb,amsthm}
\usepackage{cite,soul}
\usepackage{tikz}
\usepackage[capitalize]{cleveref}
\usepackage[margin=2.232cm]{geometry}

\usepackage{pgfplots}
\usepackage{import}
\usepackage{float}
\definecolor{ITgreen}{rgb}{0,0.8,0}
\definecolor{darkred}{rgb}{0.8,0,0.2}
\definecolor{darkblue}{rgb}{0.13,0.3,0.57}
\definecolor{darkgreen}{rgb}{0,0.5,0}
\definecolor{bluegreen}{rgb}{0,1,1}
\usetikzlibrary{arrows,calc,through,backgrounds,matrix,decorations.pathmorphing}
\usepackage{mathtools}
\usetikzlibrary{arrows}
\usetikzlibrary{decorations.pathreplacing}
\usetikzlibrary{pgfplots.groupplots}
\pgfplotsset{compat=1.16}
\usepackage{braket}
\pgfplotsset{compat=1.9}
\usepackage{hyperref}

\DeclareMathOperator{\Tri}{Tri}
\begin{document}
\title{Information Rates with Non Ideal Photon\\[-1.25ex] Detectors in Time-Entanglement Based QKD}
\author{
        Dunbar~Birnie~IV, 
        Christopher~Cheng,
        and~Emina~Soljanin,~\IEEEmembership{Fellow,~IEEE}
\thanks{The authors are with the Department
of Electrical and Computer Engineering, Rutgers, the State University of New Jersey, Piscataway, NJ 08854, USA, e-mail: (see https://www.ece.rutgers.edu/emina-soljanin).}
\thanks{This  research  is  based  upon  work  supported  by  the  National  Science  Foundation  under  Grant  \# FET-2007203}}

\markboth{IEEE Transactions on Communications, submitted 
July 2022}%
{}
\maketitle
\begin{abstract}
We develop new methods of quantifying the impact of photon detector imperfections on achievable secret key rates in Time-Entanglement based Quantum Key Distribution (QKD). We address photon detection timing jitter, detector downtime, and photon dark counts and show how each may decrease the maximum achievable secret key rate in different ways. We begin with a standard Discrete Memoryless Channel (DMC) model to get a good bound on the mutual information lost due to the timing jitter, then introduce a novel Markov Chain (MC) based model to characterize the effect of detector downtime and show how it introduces memory to the key generation process. Finally, we propose a new method of including dark counts in the analysis that shows how dark counts can be especially detrimental when using the common Pulse Position Modulation (PPM) for key generation. Our results show that these three imperfections can significantly reduce the achievable secret key rate when using PPM for QKD. Additionally, one of our main results is providing tooling for experimentalists to predict their systems' achievable secret key rate given the detector specifications.
\end{abstract}


%
\IEEEpeerreviewmaketitle


\section{Introduction}
%
%
%
%

Quantum Key Distribution (QKD) generates and distributes secret classical encryption keys between two or more users at different locations. 
We consider QKD protocols where two parties, Alice and Bob, establish a secret key by communicating over a quantum and a classical channel, which can be accessed by an eavesdropper, Eve. The quantum channel is essential in QKD for preventing undetected eavesdropping and generating randomness when using entanglement-based QKD.
At a high level, there are two main QKD steps. In the first step, Alice and Bob generate  \textit{raw key} bits by using a quantum channel. 
Their respective raw keys may disagree at some positions, could be partly known to Eve, and may not be uniformly random. In the second step, Alice and Bob process the raw key to establish a shared {\it secret key}. They communicate through the public classical channel to reconcile differences between their raw keys, amplify the privacy of the key concerning Eve's knowledge, and compress their sequences to achieve uniform randomness. 
At the end of the protocol, Alice and Bob 1) have identical uniformly random sequences and 2) are confident the shared sequence is known only to them. Therefore the secret key is private and hard to guess. For an in-depth survey of the most prominent QKD protocols, we refer the reader to \cite{qkdsurvey21}.

One of the main challenges QKD protocols face today is approaching the secret key generation rates supported by AES and other standard classical key distribution methods \cite{Diamanti2016}. Primarily, this comes down to an inability to extract all the information communicated over the quantum channel. 
Initially, QKD protocols relied on polarization-entangled photons \cite{QKD:ekert91}. Alice and Bob could extract a single key bit by measuring the polarization of entangled photon pairs. Thus, such protocols restrict the maximum secret key rate to the maximum entangled photon pair generation rate. Due to the difficulty of generating entangled photon pairs, these experiments operate in what is referred to as photon-starved conditions. To exceed this limit and to offer high key rates in photon starved conditions, we need some way of extracting multiple bits from each photon pair.  

Time-entanglement QKD promises to increase the secret key rate and distribution distances compared to other entanglement-based QKD implementations \cite{Zhong15}. 
In time-entanglement QKD schemes, an independent source (or one of the participants) randomly generates entangled photon pairs. Entangled photon inter-generation times are independent and identically exponentially distributed, giving a source of perfect randomness. Moreover, we theoretically could get arbitrarily many bits from a single photon arrival time given detectors with sufficiently precise measurement. However, such sensors do not exist. 

Alice and Bob extract the raw key bits from the arrival times of entangled photons through time binning. Each of them individually discretizes their timeline into \textit{time bins} and groups adjacent time bins into \textit{time frames}. They record photon arrivals as occupied bins within frames. They then use the position of the occupied bins within a time frame to generate random bits. The bit extraction scheme may follow pulse position modulation or some recently proposed adaptive strategies \cite{QKD:Zhou13, QKD:KarimiSW20}.
Under ideal conditions, photon inter-arrival times (as their inter-generation times) are independent and identically exponentially distributed. Several groups studied such systems, most notably \cite{Zhong15} who constructed an end-to-end, high dimensional time-entanglement experiment. 

However, non-ideal detectors suffer from jitter, dark counts, and downtime. Jitter occurs because of imprecision in the time tagging, which causes discrepancies between Alice's and Bob's raw keys. Dark counts occur due to light leakage into the system and also cause discrepancies between the raw keys. These errors reduce the secret key rate by increasing Alice and Bob's public exchange of information for key reconciliation. Downtime is the time following a photon detection during which no other detection can occur. Thus the perceived arrivals are no longer independent. In their experiment \cite{Zhong15}, Zhong et al.\ used a 50:50 beam-splitter to distribute the photon arrivals to two detectors at each station to overcome the loss of photon detections caused by downtime. These dependencies reduce the secret key rate and require Alice and Bob to compress their sequences to achieve uniform randomness. 

The adverse effects of various detector imperfections in QKD protocols have been recognized. A recent survey paper \cite{QKD:realistic} extensively studies secure quantum key distribution with realistic devices in the context of prepare-and-measure protocols, such as BB84 \cite{QKD:BennettB84}. We focus on time entanglement-based QKD. In particular, we model detector jitter, downtime, and dark counts and show how these imperfections affect the secret key rates.

This paper is organized as follows: In Sec.~\ref{sec:sysmodel}, we introduce a typical experimental time-entanglement QKD setup. We then establish a system model and describe the various imperfections we will analyze. In Sec.~\ref{sec:inforec}, we present the methods used to calculate key rate and show the trade-off between increasing raw key rate and increasing errors due to detector jitter. In Sec.~\ref{sec:downtime}, we introduce the novel Markov Chain analysis of detector downtime and use these models to investigate the severity of downtime on secret key generation rates. In Sec.~\ref{sec:darkcounts} we pivot and investigate the dark counts' effect on key generation rates. Finally, in Sec.~\ref{sec:conclusions}, we briefly summarise our results and observations. 

\section{Time-Entanglement QKD System Model}
\label{sec:sysmodel}

\subsection{Sources of Entangled Photons\label{sec:source}}
\begin{figure}[H]
  	\begin{center}			
  	    \includegraphics[scale=0.99]{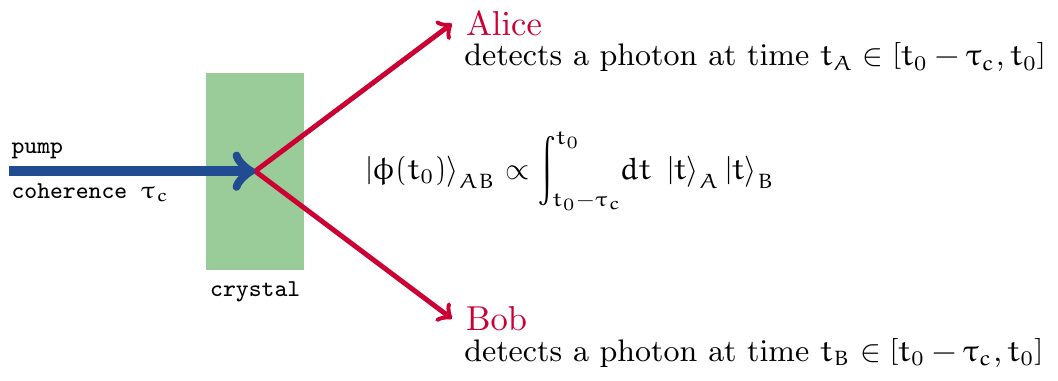}
	\end{center}
    \caption{Generation of time-entangled photons.
    Ideally, Alice and Bob receive their individual photons at identical and uniformly random times, but that is not true when they use practical, imperfect detectors.}
    \label{fig:spdc}
\end{figure}

Entangled photons are commonly generated by Spontaneous Parametric Down-Conversion (SPDC). This process occurs when a non-linear crystal is irradiated with a laser, referred to as the pump. The crystal may split the pump photon, resulting in two new photons. This event occurs randomly and relatively rarely at a rate on the order of $10^6 s^{-1} mW^{-1}$ \cite{Schneeloch_2019}. The emission of a photon pair from such sources is equally likely to occur anywhere within a window equal to the pump coherence time $\tau_c$, and it does not depend on the previous emissions. Because generation times are uniformly random and independent, photon pairs are generated according to a Poisson Process with mean $\lambda_p$. Therefore the photon inter-arrival times are exponentially distributed with the rate $\lambda_p$. This parameter can be regulated by varying the energy of the pump. Fig.~\ref{fig:spdc} shows a simplified depiction of an SPDC source.
One photon is sent to Alice, and one to Bob. The entangled state of the emitted pair is given by 
\[
\displaystyle{\ket{\phi(t_0)}_{AB}\propto\int_{t_0-\tau_c}^{t_0}\!\! dt~ \ket{t}_A\ket{t}_B}
\]
Observe that the state represents 
a uniform superposition in time within the pump coherence time $\tau_c$ from the moment $t_0$ when the SPDC photons leave the crystal. This process is 1) our source of randomness and 2) the method that ensures Alice and Bob base their keys on correlated information. 

\subsection{Single Photon Detectors}
\label{sec:spdm}
\begin{figure}[hbt]
    \centering
    \includegraphics[scale=0.8]{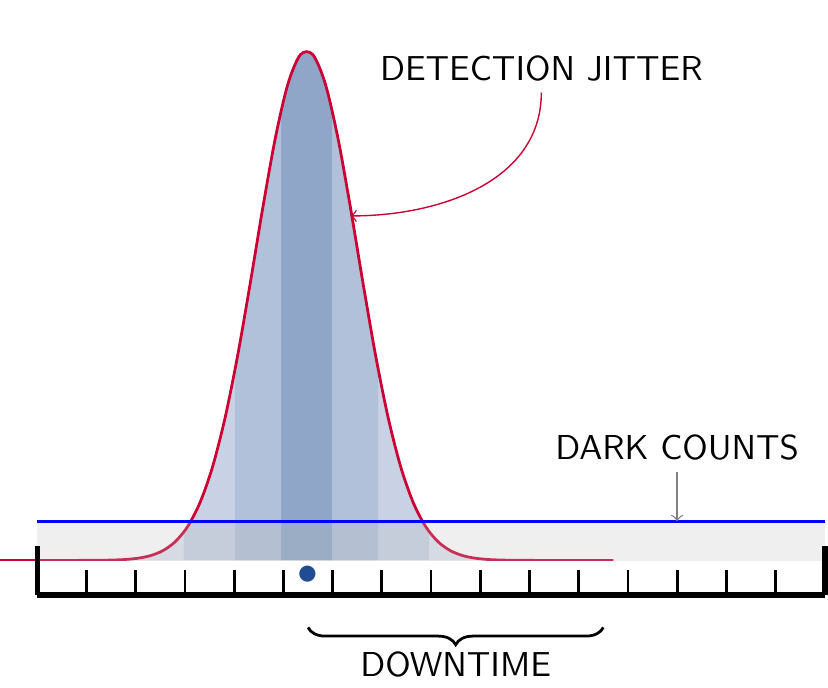}
    \caption{Detector imperfections include jitter errors, downtime, and dark counts. The Gaussian curve sketches the PDF of the detector jitter. Multiple bins may be affected, each with likelihood proportional to the area under the curve above it. Dark counts occur uniformly within the frame and here are represented by the uniform distribution.}
    \label{fig:jitteroverbins}
\end{figure}

The most common single-photon detectors are Superconducting Nanowire Single-Photon Detectors (SNSPDs). These detectors currently exhibit properties closest to those of the ideal sensors. They have high efficiency, meaning they detect the majority of incident photon arrivals accurately. They have low dark count rates, meaning they rarely report a photon detection without a photon arrival. Furthermore, they have low detector downtime $d$ and slight detector timing jitter that manifests as Gaussian noise with zero mean and variance $\sigma_d^2$. Unfortunately, these effects are non-negligible: 1) detector jitters and dark counts cause disagreements between Alice's and Bob's keys, and 2) the downtime introduces memory within the raw key bits. This paper focuses on the secret key rate loss due to these non-ideal properties. Fig.~\ref{fig:jitteroverbins} illustrates three of these detector imperfections.

Let $t_A$ be the time when one of the photons of the pair generated by the source is registered by Alice's detector, and $t_B$ be the time the other photon is registered by Bob's detector. Then
\begin{equation}
 t_A = u+\eta_A ~\text{and} ~ t_B = u + \eta_B.   
\label{eq:co}
\end{equation}
Here $\eta_A$ and $\eta_B$ are independent samples of a Gaussian distribution with the zero mean and variance $\sigma_d^2$, and $u$ is a sample of a random variable $U$ distributed over the interval $[t_0-\tau_c,t_0]$. We take $u$ to be the actual arrival time of the SPDC photon.

Let $T_A$ and $T_B$ be the random variable associated with the times Alice and Bob register their photon arrivals. The jitter errors cause the discrepancy between $T_A$ and $T_B$ to be distributed according to the convolution of the individual detector jitter probability density functions, which corresponds to a Gaussian distribution with twice the variance of each detector jitter. Therefore
\[
T_B-T_A \sim N(0, 2\sigma_d^2).
\]

Dark counts arrive uniformly, and independently, at either detector and are indistinguishable from the SPDC photons. This can cause significant errors if they make up a large fraction of the detected photons. Dark counts are primarily a result of light leakage into optical lines.

When an SPD detects a photon, it enters a temporary state wherein no other photon arrivals can be detected. This means a possibility of our previously mentioned interval (which was $[t_0-\tau_c,t_0]$) being truncated or eliminated entirely. This effect further limits the observed photon arrival rate.

\cite{SNSPD:Zadeh2021}

\subsection{Eavesdropping Model}
The described system allows raw key generation at two locations. Since there is no passive eavesdropping possible on a quantum channel, Alice and Bob can always detect the presence of Eve. Systems implementing this kind of QKD experiment have been recently shown to achieve photon information efficiency up to 4.082 secure-key bits/photon and a secure-key rate up to 237-kbit/s \cite{9572792}. They commonly pass a fraction of photons through a special interpreter to produce entangled photons in the maximally entangled state
$\ket{\varphi_{AB}}\propto\ket{0_A0_B}+\ket{1_A1_B}$. Alice and Bob can quantify Eve's information gain based on such photons by playing a variant of the CHSH game. We assume that they halt the key distribution if they detect an eavesdropper beyond the non-classical bound of the CHSH game. Thus, the raw keys that Alice and Bob receive have guaranteed security against eavesdropping on the quantum channel. Equivalently, we could say there has been no eavesdropping on the quantum channel. We focus on the second phase of the key distribution process wherein Alice and Bob take their arrival sequences and map them into identical secret keys. 

Due to the detector imperfections,  Alice's and Bob's raw keys do not entirely agree (jitters and dark counts cause errors). They are not uniformly random either (detector downtime introduces memory). Therefore, Alice and Bob must first perform information reconciliation to establish key agreement. This process involves transmission over the public channel. Alice and Bob must then perform privacy amplification to regain the lost randomness due to downtime and recover the secrecy lost in the information reconciliation process. This ultimately reduces the key rate such that both these goals may be met. This paper focuses on computing information-theoretic bounds on the key rate loss in the privacy amplification step. In practice, information reconciliation is carried out by using error-correcting codes whose rate will ultimately determine the secret key rate. Practical information reconciliation error-correcting codes for PPM schemes have been recently proposed in \cite{QKD-C:YangSCWD19, QKD-C:boutrosS22}.

\subsection{Raw Key Extraction}
Alice and Bob rely on the correlated random photon arrivals to generate their secret keys. There are many ways to extract keys from this correlated information. One popular method is similar to Pulse Position Modulation (PPM); see, e.g., \cite{Zhong15} and references therein. In PPM, Alice and Bob synchronize their clocks and discretize their timelines into time frames of size $T_f$ each consisting of $n$ time bins. Each time bin has a width of $\tau_b = T_f/n$. 
In PPM, Alice and Bob agree to retain only time frames in which only one bin is occupied, while discarding all other frames. Individual photons are considered to occupy a time bin depending on where within the frame it arrives -- if a photon arrives at time $t_A$, the occupied frame is frame number $\lfloor t_A / T_f \rfloor$ and the PPM decoding of this frame is $\lfloor (t_A\mod T_f) / \tau_b \rfloor$. 

The number of raw key bits $K_{\text{raw}}$ that PPM decoding can extract from each retained frame is
\begin{equation}
 K_{\text{raw}}  = \log_2 (n) ~ \text{bits per frame.}
\label{eq:RawKey}   
\end{equation}
To find the overall number of raw key bits extracted on average $R_{\text{raw}}$, we need to know the probability that a frame is valid, and for that, we need to know the probability that a bin is occupied. Since photon inter-arrival times are exponentially distributed (see Sec.~\ref{sec:source}), the number of photons that arrive in a single bin is Poisson distributed. From this, we can derive the probability $p$ of a bin being occupied, 
\begin{equation}
    p=1-\exp(-\lambda_p \tau_b).
    \label{eq:BinFillProb}
\end{equation}
For now we assume that the probability of each bin being occupied is independent of other bins. We can thus treat the occupancy of each bin as being modeled by a Bernoulli distribution, with parameter $p$. Hence, the probability of a frame being retained is the probability that only a single bin is occupied, which is $np(1-p)^{n-1}$. Therefore, the highest key rate achievable when using PPM is
\begin{equation}
    R_{\text{raw}} = K_{\text{raw}}np(1-p)^{n-1} = \log_2(n)np(1-p)^{n-1} ~ \text{bits per frame.}
    \label{eq:RrawPPMBest}
\end{equation}

\begin{figure}[hbt]
    \centering
    \includegraphics[]{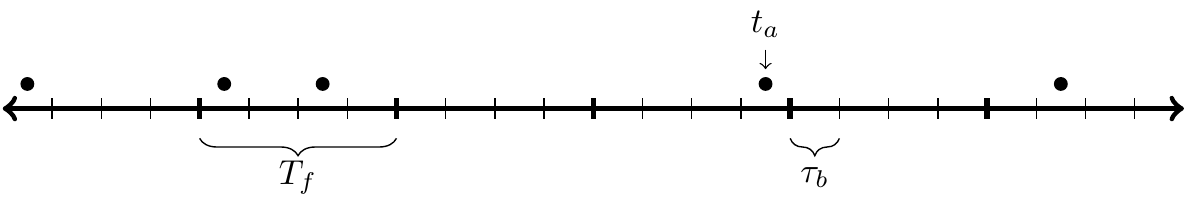}
    \caption{Time frames of duration $T_f$ consist of $n=4$ time bins. Each bin has the width $\tau_b=T_f/n$ Photon arrivals are shown as points above the line. Each is said to "occupy" the bin it lands within.}
    \label{fig:time_line}
\end{figure}

Fig.~\ref{fig:time_line} illustrates the described time of arrival approach to key generation.
In this example, each frame has four bins. The first frame is valid for PPM because there is only a single occupied bin. The second, third, and fifth frames are invalid because they have two, zero, and zero arrivals, respectively. The fourth and the sixth frames also have only one arrival, and so are valid for PPM. Here, we extract $\log_2 4=2$ bits per each PPM valid (retained) frame as follows: The first frame has only the first bin occupied, which renders bits $00$. Frame four would give $11$, and frame six, $01$. 

\begin{figure}[hbt]
    \centering
    \includegraphics[]{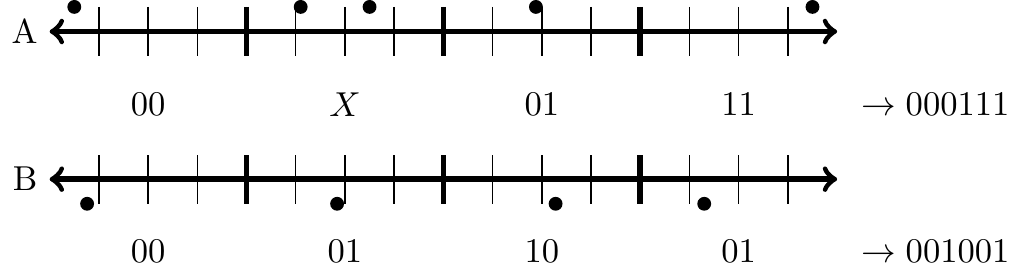}
    \caption{Two potential arrival sequences at Alice's and Bob's stations. PPM decoding shown below each frame and the resulting key sequences at the right.}
    \label{fig:twotimelines}
\end{figure}

To illustrate potential disagreements between Alice's and Bob's extracted raw key bits, we consider an example in Fig.~\ref{fig:twotimelines}. Alice and Bob had discretized their timelines into time frames consisting of four time-bins.
We can see that in the first frame bit extraction works as expected, and both parties extract the same key bits despite some slight timing jitter that stays within the time bin. However, they both throw out the second frame because Alice detected an extra dark count in her frame. The third time frame shows how the timing jitter can cause errors between the extracted keys. Here the jitter caused the photons to be detected in separate bins and in this case caused two bitwise errors in the final key sequences. The fourth frame shows an example where Alice and Bob detect a dark count in the same time frame. Before information reconciliation, these seem like valid PPM frames despite the resultant bits being entirely uncorrelated. The rates of these two kinds of errors are significantly affected by the specific design parameter choice. For example, smaller bins lead to more potential bits per time frame but increase the likelihood of jitter errors, while larger time frames can lead to a greater number of unusable time frames due to multiphoton arrival events (as in the second pair of frames in Fig.~\ref{fig:twotimelines}.) 

\begin{figure}[hbt]
    \centering
    \includegraphics[]{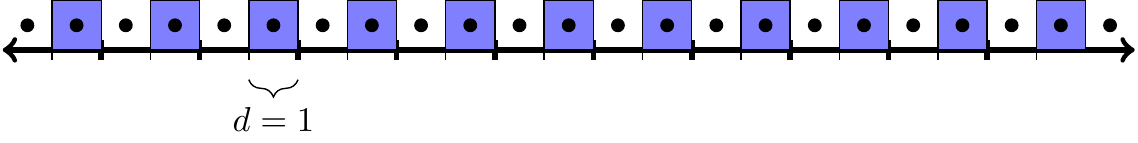}
    \caption{An example of a deterministic sequence that appears acceptable to Alice and Bob because of the detector downtime. Each time frame consists of $n=2$ time bins, and the photon generation probability is high. Photon arrivals are shown as points above the line. Every bin is occupied, but the detector downtime (shown in blue) occludes every other arrival tricking Alice and Bob into accepting and decoding each time frame.}
    \label{fig:downtimeppm}
\end{figure}

Finally, Fig.~\ref{fig:downtimeppm} illustrates the impact of detector down time on key extraction. 
In this example, we have naively parameterized PPM so that the time bin duration equals the detector downtime. We assume that the photon arrival rate is sufficiently high to ensure each bin will detect a photon arrival with a high probability. The resulting key sequence is the all-zero string and is entirely predictable (i.e., carries no randomness). This particular case is contrived and we can avoid such scenarios, but as we'll see, the entropy reduction phenomenon ultimately persists albeit to a lower degree.

\section{Paper Goals and Summary of Results}
We are now ready to more precisely state this paper's goals and summarise the results. 
This paper has three goals addressed in the next three sections. We briefly describe them below.

\subsection{Information Loss Due to Detector Jitter}
Alice and Bob get their raw key bits based on the correlated (ideally coincidental) arrival times of the SPDC-generated entangled photon pairs. The level of correlation is determined by the detector jitter noise variance and the time bin size (since we extract the raw key bits by time binning). The smaller the time bin, the higher the raw key rate. On the other hand, the smaller the time bin, the more Alice's and Bob's raw keys may disagree due to jitter.
An increase in the disagreement between their keys necessitates an increase in the public communication for key reconciliation, thus, reducing the secret key length. 
We compute the number of bits $\beta_r$ that have to be sent from Alice to Bob to reconcile their raw keys as a function of the jitter variance and bin size. We observe that when the detector jitter variance increases, the bin count that maximizes the post reconciled key rate decreases. These results and trade-offs are shown in Sec.~\ref{sec:inforec}.

\subsection{Information Loss Due to Detector Downtime}
When the detector exhibits downtime, Alice's and Bob's raw key bits are not uniformly random, as we illustrated in the exaggerated example of Fig.~\ref{fig:downtimeppm}. Alice and Bob have to compress their reconciled raw keys to achieve uniform randomness. We compress the keys by the compression rate $C^d_r$, which is a function of the detector downtime, photon generation rate, and the number of bins per frame. To characterize the impact of detector downtime on the system, we model combined detector and time binning operations by Markov Chains (MC). The entropy rate of the system's MC determines the minimum compression rate to guarantee the key's uniform randomness. These MCs are complex, and we developed an algorithm to create them for various parameter values, implemented in an online tool available at \href{https://cc1539.github.io/qkd-binning-demo-2}{https://cc1539.github.io/qkd-binning-demo-2}. Sec.~\ref{sec:downtime} presents these results and shows how the compression rate $C^d_r$ depends on downtime.

\subsection{Modeling the Impact of Dark Counts}
We introduce a novel method for analyzing the impact of dark counts in entanglement QKD using PPM. We develop a way to model coincident dark counts indistinguishable from valid PPM frames. In doing so, we characterize the severity of dark counts in PPM more accurately than in prior efforts \cite{Brougham16}. This work is shown in Sec.~\ref{sec:darkcounts}.

\subsection{Estimating the Secret Key Rate\label{sec:FinalEstimate}}
%

The key rate simultaneously suffers from all impairments discussed above. This paper only looks into the rate loss caused by the detector jitters and downtime and provides and proposes a model for dark counts. We address the detector jitter noise and downtime separately because these two impairments cause rate loss in very different ways. We next describe their combined effect on the key rate.

We define two key rates of interest. First, $K_{\text{reconciled}}$ corresponds to the number of key bits per PPM frame after reconciliation and privacy amplification. Then, $K_{\text{uniform}}$ corresponds to the number of key bits per PPM frame after compression to remove downtime-induced memory. 
\begin{equation}
K_{\text{reconciled}} = K_{\text{raw}} - \beta_r
~ \text{and} ~
K_{\text{uniform}} = C^d_r\cdot K_{\text{raw}}
    \label{eq:metrics}
\end{equation}
Combining the two, we obtain $K_{\text{secret}}$. This represents the number of secret key bits per PPM frame, and is given by
\begin{equation}
K_{\text{secret}} = C^d_r(K_{\text{raw}} - \beta_r).
    \label{eq:final}
\end{equation}

We have one step left to our ultimate performance metric. Recall that PPM only retains frames with a single arrival. Therefore to get the average key rate $R$, the key bits per valid frame $K$ have to be multiplied by the probability that a frame contains exactly one occupied bin, which is $np(1-p)^{n-1}$. That is,
\begin{equation}
    R_{\text{secret}} = np(1-p)^{n-1} K_{\text{secret}}
\end{equation}

The following sections will show how to separately evaluate the rate-loss metrics in equation \eqref{eq:final} that computes their joint influence on the secret key rate. We do not look into different impairments simultaneously in the following sense.
In evaluating the downtime effect in Sec.~\ref{sec:downtime}, we disregard jitter errors. Therefore, only downtime introduces memory. In evaluating the jitter effect in Sec.~\ref{sec:inforec}, we assume that the difference between Alice's and Bob's photon detection is the difference between two zero-mean Gaussian random variables. Instead, if we took the influence of the downtime simultaneously, the mean of these Gaussian random variables would also be a random variable whose mean is zero because of the symmetric jitter, which could have caused Alice's downtime to start before or after Bob's. Thus our jitter analysis is the expected case analysis. More detail is provided in the corresponding sections below.

\subsection{Notation and Assumptions}
As a reminder and a reference, we list the previously defined system parameters and variables in Table~\ref{table_of_params}.
The table also lists some relationships between the parameters.
\begin{table}[hbt]
  \caption{
  QKD System Parameters and Variables}
    \label{table_of_params}
    \centering
    \begin{tabular}{r c l}
        $\lambda_p$ & -- & entangled photon generation rate\\
        $d$ &  -- & detector downtime\\
        $\sigma_d^2$ &  -- & detector jitter variance\\
        $T_f$ &  -- & PPM frame width\\
        $n$ &  -- & number of time bins per frame\\
        $\tau_b$ &  -- & time bin width, $\tau_b = T_f/n$\\ 
        $p$ &  -- & bin occupancy probability, $p=1-\exp(-\lambda_p \tau_b)$\\
        $K_{\text{raw}}$ & -- & number of raw key bits per PPM frame,
        $K_{\text{raw}}  = \log_2 n$ \\
        $\beta_r$ & -- &number of bits sent over the public channel for reconciliation\\
        $C^d_r$ &  -- & downtime compression ratio \\ 
    \end{tabular}
\end{table}
This paper assumes that $T_f$, $\lambda_p$, and $d$ are system parameters determined by a particular system setup and equipment. The design parameter we can choose in an attempt to optimize performance metrics is the number of bins per frame, $n$.

We adopt models widely used in the literature, e.g., Poisson photon statistics (described in textbooks such as \cite[Chapter 5]{Optics:Fox}) and Gaussian jitters which are commonly denoted in terms of full width half maximum (FWHM)\cite{gaussianJitter} (and described this way in detector specification sheets such as \cite{PhotonSpot}). For further experimental details and device specifications we direct the reader to \cite{Zhong15} and \cite{SNSPD:Zadeh2021}. However, our theoretical approach, derivations, and numerical evaluations do not rely on such assumptions. The parameter values used in our numerical examples are selected based on the corresponding instances in the literature and to emphasize specific points.

\section{Information Loss due to Detector Jitter}
\label{sec:inforec}
\subsection{The Source Model Secret Key Agreement Rate}
The secret key rate is the "maximum rate at which Alice and Bob can agree on a secret key while keeping the rate at which Eve obtains information arbitrarily small" \cite{Maurer93a}. In the case of time-entanglement-based QKD, Alice and Bob obtain correlated streams of bits (raw keys) based on their respective Time of Arrival (ToA) measurements, as described in the previous section. However, they must communicate to agree on a key, i.e., reconcile their differences. Every communication required for this process must be considered open communication accessible to Eve. Here, we focus on one-way information reconciliation schemes in which Alice sends information about her sequence to Bob, who uses it to correct the differences between his and Alice's raw keys. 

After one-way information reconciliation, Alice and Bob share Alice's initial raw key. However, since they communicated over a public channel, the shared key is not secret. To correct that, Alice and Bob perform privacy amplification. They commonly hash their shared keys, establishing secrecy but shortening the key. The secret key rate is the post-privacy amplification key length per second. This section aims to characterize the secret key rate when Alice and Bob obtain their raw key rates in the presence of detector jitters. Recall that in this case, Alice and Bob base their secret key generation on correlated arrival times given in \eqref{eq:co}.
We know that, for this model (referred to as the source model in Information Theory 
\cite[Chapter~22.3]{books:GK2011}), the secrecy capacity is equal to the mutual information between Alice's detected arrival time $T_A$ and Bob's detected arrival time $T_B$,
\begin{equation}
    C_K = I(T_A; T_B),
    \label{eq:scc}
\end{equation}
when the eavesdropper has access to the public communication but does not have correlated prior information (see, e.g., \cite[p.~567]{books:GK2011}). 

\subsection{Secret Key Rates in PPM Key Extraction}
Let us consider the case where Alice and Bob use the PPM scheme to extract raw key bits, and jitter errors are present as described in Sec.~\ref{sec:spdm} with variance $\sigma_d$. The two observe discrete correlated random variables identifying the occupied bin in the frame. When the number of bins per frame is $n$, Alice observes $X_n$ and Bob observes $Y_n$ given by
\begin{equation}
    X_n= U_n+J_{A,n}~\text{and}~Y_n= U_n+J_{B,n}
    \label{eq:dcm}
\end{equation}
where $U_n$ is uniform over the set $\{0,1,\dots,n-1\}$, and $J_{A,n}$ and $J_{B,n}$ have integer support, taking value $k$ with the probability 
\begin{align}
 P\bigl[J_{A,n} = k\bigr] & =\frac{1}{\sqrt{2\pi\sigma_d^2}}\int_{(k-1/2)\tau_b}^{(k+1/2)\tau_b}\exp(-t^2/(2\sigma_d^2))\,dt.\nonumber\\
  & = \frac{1}{\sqrt{2\pi}}\int_{(k-1/2)T_f/(n\sigma_d)}^{(k+1/2)T_f/(n\sigma_d)}\exp(-\tau^2/2)\,d\tau.
\label{eq:prob}   
\end{align}
Observe that because of the jitter in the previous frame, the downtime will cause a difference in $U_n$ for Alice and Bob. However, since the jitters are symmetric, this difference will have a zero mean.

The rate of disagreement (RoD) between Alice's and Bob's raw keys is given by  
\begin{equation}
    \text{RoD} = 1 - P\bigl[J_{A,n}=J_{B,n}\bigr]\\
    \label{eq:ser}  
\end{equation}

\begin{figure}[hbt]
\centering
 \begin{tikzpicture}
 \node[left] at (0,0) {\includegraphics[scale=0.5]{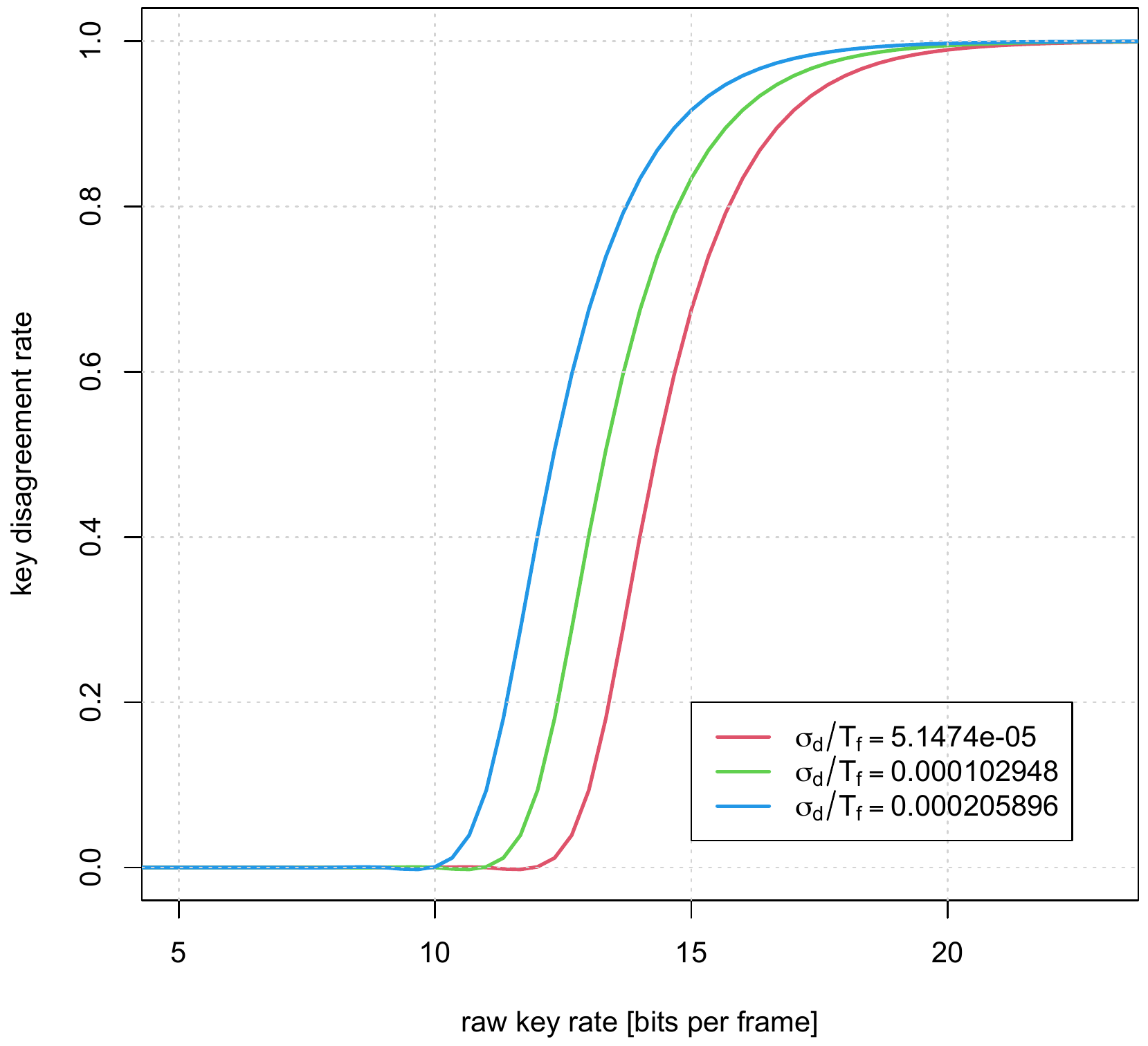}};
 \node[right] at (0,0) {\includegraphics[scale=0.5]{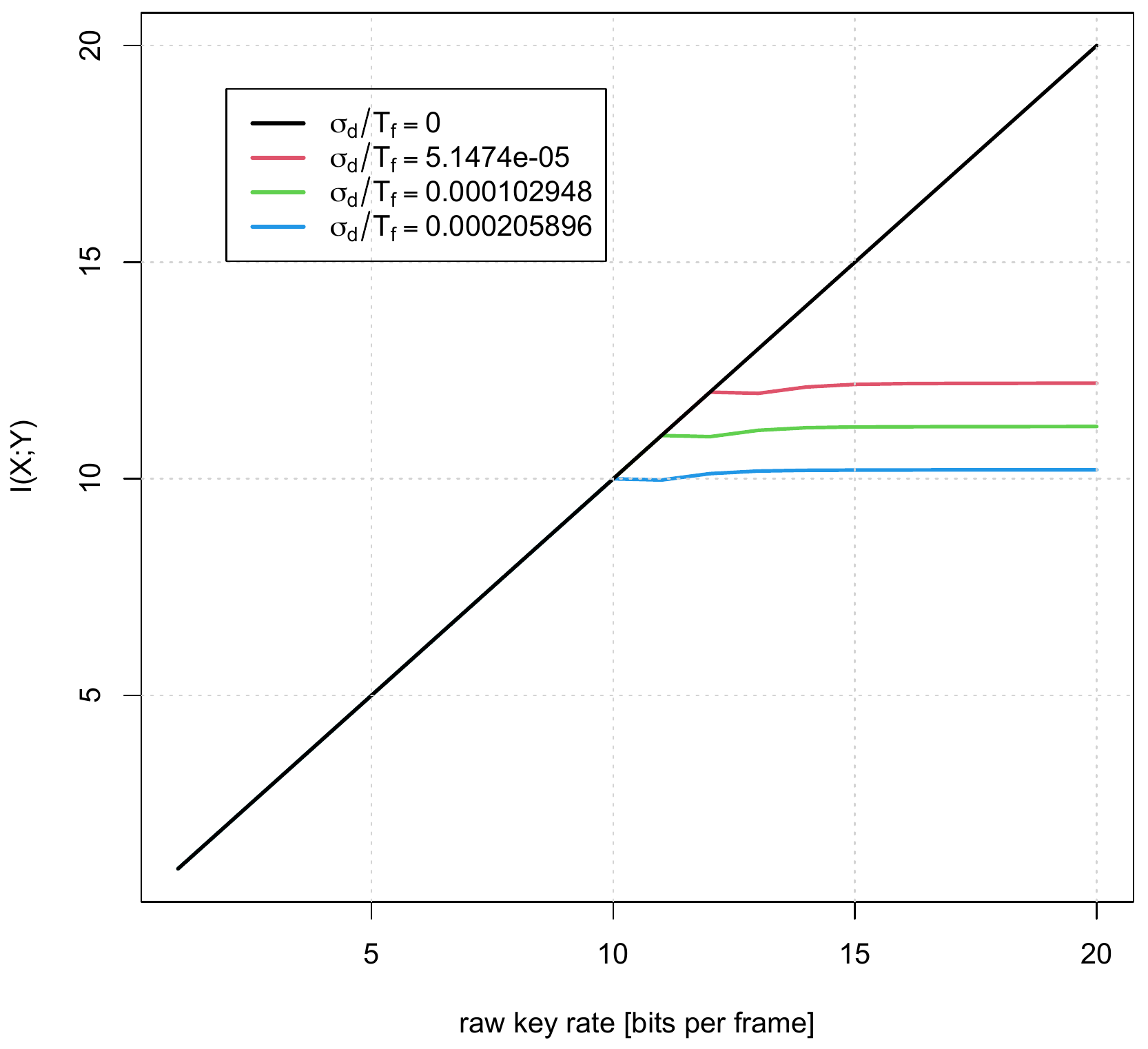}};
 \node at (-8.5,4) {(a)};
  \node[right] at (0,4) {(b)};
 \end{tikzpicture}
    \caption{(a) Dependence of the disagreement rate between Alice's and Bob's keys on the raw key rate (or equivalently, the bin size), cf.~\eqref{eq:ser}. As bin size decreases, both the raw key and disagreement rates increase. (b) Dependence of the maximum reconciled key rate on the raw key rate, cf.~\eqref{eq:skr}.
    A larger raw key disagreement leads to more reconciliation information being revealed through the information reconciliation process.}
    \label{fig:postreconcilliationrates}
\end{figure}

Observe from \eqref{eq:prob} and \eqref{eq:ser}, that the key-bit disagreement happens at a rate that depends on the ratio of the detector noise variance to time frame ratio $\sigma_d/T_f$ and the number of bins per frame $n$. The PPM time frame size is constrained by the pump coherence time. The reported coherence time in \cite{Zhong15} is 250-330 ns. Recall that the number of bits extracted per PPM-valid frame with the single occupied bin out of $n$ in the frame is equal to $\log_2 n$. Thus, increasing $n$ gives more raw key bits per frame. However, the bin width decreases as $n$ increases, making the binning process more prone to errors; 
See Fig.~\ref{fig:jitteroverbins}.
In Fig.~\ref{fig:postreconcilliationrates}a, we vary the number of bins per frame and observe the key disagreement rates. The values for $\sigma_d/T_f$ are selected by choosing the equivalent ratio for an SNSPD with FWHM jitter of 80ps or $\sigma_d = 33.97 ps$, and $T_f = 330 ns$ as in \cite{Zhong15}. 

The secrecy capacity of our PPM scheme for the number of bins equal to $n$ is 
\begin{equation}
    C_{K,n} = I(X_n;Y_n),
    \label{eq:scd}
\end{equation}
where $X_n$ and $Y_n$ are given by \eqref{eq:dcm}. We say that $I(X,Y) = \lim_{n\to\infty} I(X_n;Y_n)$ is the ultimate achievable secret key rate.

To compute $I(X_n;Y_n)$, we observe that Bob perceives Alice's detection as corrupted by the difference of the two independent Gaussian random variables (Alice's and Bob's noise). We can therefore compute the transition probabilities $p(y|x)$ of the event that Bob detects a photon arrival in bin $y$ given that Alice detects her entangled photon arrival in bin $x$. We calculate these transitions by integrating the double-jitter error distribution over the domain of the specific bin where Bob has detected his arrival, as 
\begin{equation}
    p(y|x)=
    \frac{1}{\sqrt{4\pi\sigma_d^2}}\int_{(y-x-1/2)\tau_b}^{(y-x+1/2)\tau_b}
    \exp\bigl[-t^2/(4\sigma_d^2)\bigr]\,dt.
    \label{eq:tp}
\end{equation}
We note that for a given $x$ the transition probabilities are a function of the difference $x-y$. This phenomenon allows us to approximate the mutual information in \eqref{eq:scd} as follows
(see \cite[Ch.~7.2]{books:CT2006}):
\begin{align}
    K_{\text{reconciled}} & < {C}_{K,n} = I(X_n;Y_n) = \log_2 n - H(Y_n|x)   \label{eq:skr}\\
    & = K_{\text{raw}} - H(Y_n|x) ~ \text{bits per frame,}\nonumber
\end{align}
where $H(Y_n|x)$ does not depend on $x$ and can be computed based on the transition probabilities defined in \eqref{eq:tp}. 
Expression \eqref{eq:skr} represents an upper bound on the number of secret key bits per frame after information reconciliation. The parameter $\beta_r = K_{\text{raw}} - K_{\text{reconciled}}$ (see \eqref{eq:metrics} above) depends on the rate of the error correcting code used for reconciliation. The value of $\beta_r$ depends on the efficiency of the chosen code. A more thorough discussion on key rates, other ways to derive them, and coding is given in \cite{QKD-C:boutrosS22}.

Fig.~\ref{fig:postreconcilliationrates}b plots ${C}_{K,n}$ for three different values of $\sigma_d/T_f$ selected as described above for the disagreement rate plot in Fig.~\ref{fig:postreconcilliationrates}a. We see that there is hardly any increase in the maximum achievable secret rate beyond a certain number of bins per frame which depends on $\sigma_d/T_f$. Comparing the plots in the two parts of Fig.~\ref{fig:postreconcilliationrates}, we can observe that the noticeable rate increase stops almost as soon as the key disagreement begins. 

\begin{figure}[hbt]
\centering
  \includegraphics[scale=0.7]{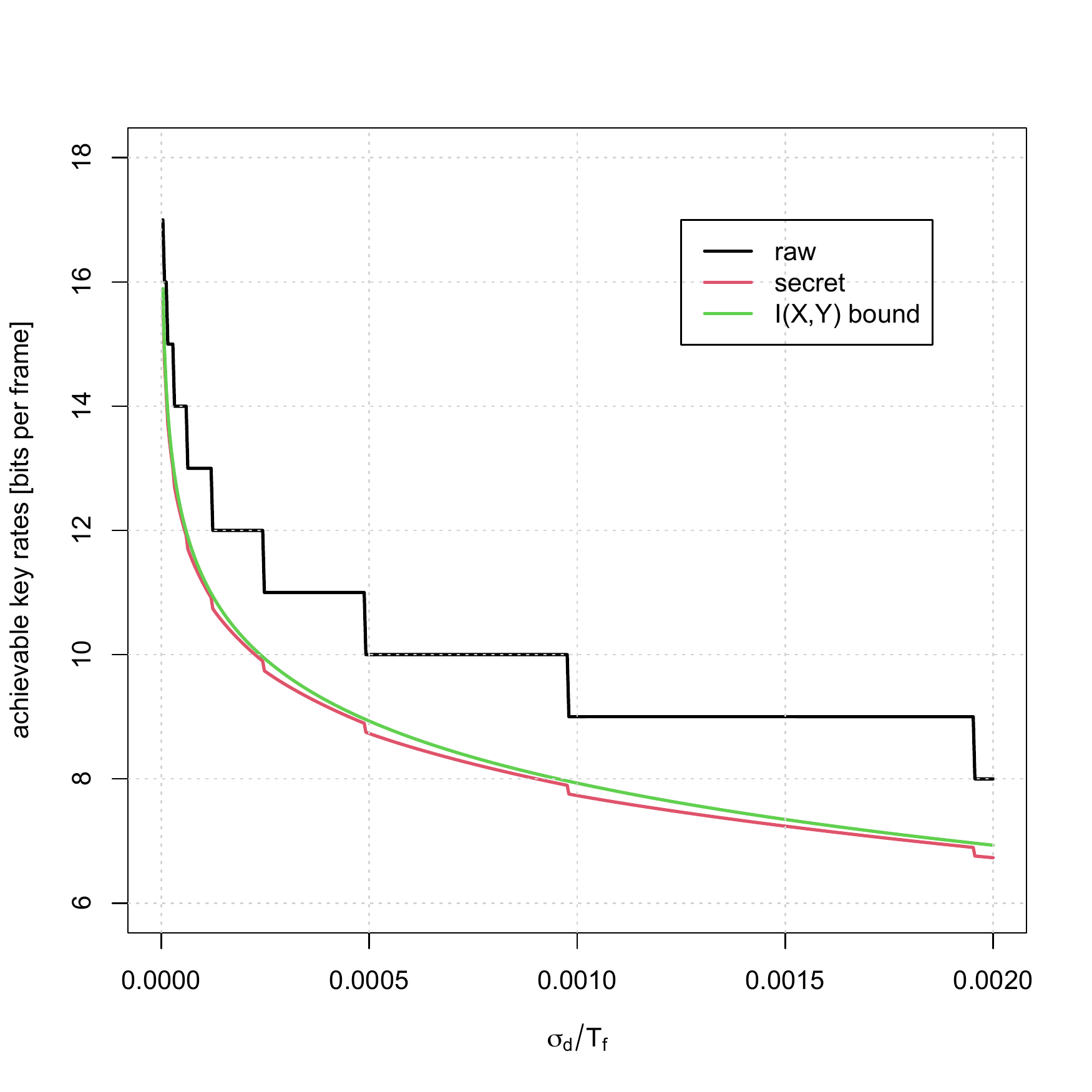}
    \caption{Dependence of the key rates on the noise level described by $\sigma_d/T_f$. The plot demonstrates that the number of raw key bits per frame determined by \eqref{eq:pgn} is sufficient for achieving a secret key rate that approaches the maximum possible.}
    \label{fig:maxIXYvssigmaratio}
\end{figure}

This observation motivated us to ask how close to the bound we can get by selecting the bin size such that it covers the interval $[-\sigma_d/T_f, +\sigma_d/T_f]$ around its central point, or equivalently,
\begin{equation}
    n = \lceil 2T_f/\sigma_d\rceil 
    \label{eq:pgn}
\end{equation}
Fig.~\ref{fig:maxIXYvssigmaratio} plots the ultimate bound $I(X,Y)$ on the reconciled key rate for unbounded $n$. We see that increasing the jitter noise variance to frame size ratio lowers the post-reconciled key rate.
Fig.~\ref{fig:maxIXYvssigmaratio} also plots the raw key rate that is given by $\log_2(n)$ where $n$ is estimated in \eqref{eq:pgn} and the secret key rate that we can achieve with that raw key rate. We see that
this secret key rate closely approximates the maximum key rate. Other approximations can be found in our related work \cite{QKD-C:boutrosS22}

Unlike previous approaches, this DMC modeling approach is straightforward and easily generalizes to any symmetric noise distribution. Earlier work considered jitters to occur within one to three time-bins \cite{Brougham16}. This work ignores the possibility of having significant jitters when using very small bin sizes, which is one of the main areas of interest. Previous approaches also force the analysis of dark counts to be considered separately. This is inconsistent with how coincident dark counts manifest in PPM. Our model accounts for all possible jitter errors, not only those close to the original bin, while simultaneously considering the effect of dark counts. We address the case of non-zero dark counts in Sec.~\ref{sec:darkcounts}, where the model is the same, but the jitter error distribution builds to account for dark counts in PPM. 

\section{Information Loss due to Detector Downtime}
\label{sec:downtime}


Under the effect of downtime, Alice's and Bob's raw keys are no longer uniformly random bits, as illustrated in the exaggerated example of Fig.~\ref{fig:downtimeppm}. Recall that our sequence of bin occupancies  (what we run through a binning scheme to extract the raw key bits) is a Bernoulli trial sequence. We have Bernoulli trials when there is no jitter or when the jitter is symmetric, which is the case here. With downtime, this is no longer the case. Photon detection in one bin prevents detection in subsequent bins, thus removing the memoryless behavior of photon arrivals. Therefore, Alice and Bob must compress their raw key sequences to obtain a uniformly random shared key.

This section focuses on computing the average number of secret key bits per frame after the compression. A family of simple Markov chains describes the dynamics of downtime-afflicted systems (see Fig.~\ref{fig:detector_markov_chain}) if we don't consider the use of PPM (or any binning scheme) to extract raw key bits. The entropy rates of these Markov chains would then determine the secret key rate. When we bring PPM back into the picture, however, the Markov chains used quickly end up with many states, complicating the computation of their entropy rates. To combat this increase in computational complexity, we introduce a family of augmented Markov chain models that can replace the original Markov chain method in certain conditions to calculate a tighter bound on the entropy rate while keeping the number of states tractable.

\subsection{The Impact of Downtime}
\begin{figure}[H]
    \centering
    \includegraphics{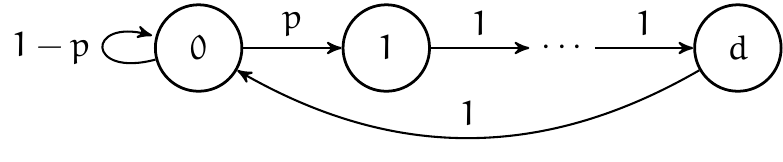}
    \caption{Possible states of detectors with downtime $d$ and possible state transitions with arrows showing the probability of transitions between each state. This Markov Chain shows that the detector capable of detecting photons sees a photon with probability $p$, and once it does, it enters into a period of downtime lasting for $d$ states.}
    \label{fig:detector_markov_chain}
\end{figure}

When we consider detector downtime behavior but ignore the PPM bit extraction, we can model the detector dynamics by the Markov Chain in Fig.~\ref{fig:detector_markov_chain}. 
The entropy rate of this gives us an upper bound on the information rate of our extracted key,
\begin{equation}
    R_{ToA}(p,d) \leq \frac{h(p)}{n(1+pd)} ~ \text{ bits/frame},
    \label{eq:basic_upper_bound}
\end{equation}
where $h(p)$ is the binary entropy. 

\begin{figure}[hbt]
    \begin{minipage}[c]{0.45\textwidth}
    \includegraphics[width=\textwidth]{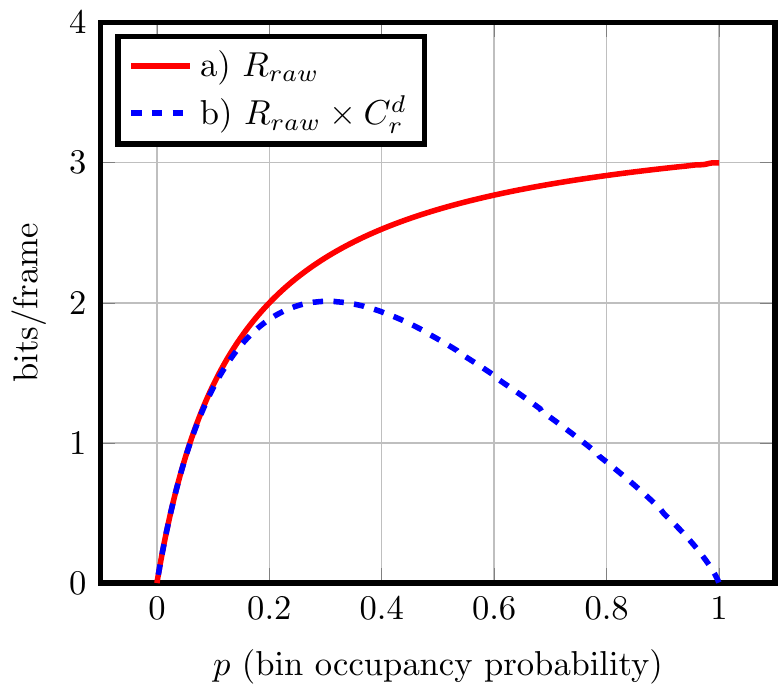}
  \end{minipage}\hfill
  \begin{minipage}[c]{0.45\textwidth}
    \caption{
       The PPM key rate as a function of $p$, the probability that a bin is occupied. The parameter $p$ is given as a function of $\lambda_p$ and $\tau_b$ in \eqref{eq:BinFillProb}. In plot a), we are \textit{not} taking into account the loss of entropy by any mechanism and only show the number of bits we get per bin as a result of using the PPM binning scheme on the input we get from the detector. Note that this is not a plot of Eq.~\ref{eq:RrawPPMBest}, which was derived under the assumption that all bins were i.i.d., an assumption that no longer holds due to downtime. In plot b), we \textit{are} taking into account the loss of entropy as a result of downtime, giving us a different value which we can call $R_{\text{raw}}\times C^d_{r}$. These plots were generated using our online tool, available at \url{https://cc1539.github.io/qkd-binning-demo-2/}
    } \label{fig:naiive_and_actual_rate}
  \end{minipage}
\end{figure}

Observe that as $d$ increases, the maximum key rate decreases. Naively, we may set the value of $p$ high, thinking that more photons will necessarily give us a higher information rate. However, that is not the case, especially for binning schemes such as PPM. To see why, consider plot a) in Fig.~\ref{fig:naiive_and_actual_rate}, which shows the key rate as a function of $p$ when we use PPM to extract key bits. 
As $p$ approaches $1$, the raw key rate seems to a maximum. However, we saw before that the information rate approaches 0 at the extremes of $p$. Indeed, when $p$ is high, the entropy of the raw key gets lower and lower until, at $p=1$, we theoretically have a deterministic sequence that may yield many bits but no randomness and, thus, no security against attackers aware of the system configuration. Consider plot b) in Fig.~\ref{fig:naiive_and_actual_rate}, which considers the loss of randomness due to downtime.

Note that if downtime is low enough compared to the frame size, this effect does not manifest, and the raw key rate changes just as the information rate does. However, in practice, values for downtime and frame size such as those provided by experiments \cite{muratComms} suggest that downtime is typically much longer than a single frame width. In this regime, we see the phenomenon explained earlier, with the raw key rate increasing while the average information extracted by each frame decreases. The behavior observed in plot b) of Fig.~\ref{fig:naiive_and_actual_rate} is thus the behavior we might expect to see in live systems. To reiterate, while increasing the raw key rate (without changing bin size) improves your key rate up to some point, one must be careful not to increase $\lambda_p$ (and subsequently $p$) too high without taking the effects of downtime into account.

\subsection{Constructing Markov Chains to Compute Information Loss}

In order to produce plot b) in Fig.~\ref{fig:naiive_and_actual_rate}, we need to expand on the idea of using Markov Chains. The Markov Chain in Fig.~\ref{fig:detector_markov_chain} is not adequate since it describes the detector on a bin-by-bin basis, while the use of binning schemes means that we must describe the detector on a frame-by-frame basis. For this, we use the techniques described in \cite{Cheng:Thesis:2022}. The first thing we do is construct what, in this context, we will call \textit{Input} Markov Chains (IMC). These represent what the detector sees, with only the additional processing step of assigning each photon detection to a bin. Analysis of this Markov Chain is sufficient to produce $R_{\text{raw}}$, which is what plot a) in Fig.~\ref{fig:naiive_and_actual_rate} shows. But to go from a) to b), we need an additional step.

This additional step is the \textit{Output} Markov Chain (OMC). Each state represents the output of a frame after applying your binning scheme of choice. For example, for PPM with $n=8$, the states of an OMC may correspond to all binary strings of length $\log_2(8)=3$. The Markov Chain Entropy of the OMC gives us $C^d_{r}\times 2^{|\text{OMC}|}$, where $|\text{OMC}|$ is the number of states in our OMC. The factor $C^d_{r}$ represents the degree to which we would need to compress the bits extracted using our binning scheme to achieve perfect randomness. For example, when $d=0$ we find that $C^d_{r}=1$ for all $p$, reflecting the fact that no compression is required as a result of downtime since there is no downtime. The adjusted key rate, $R_{\text{raw}}\times C^d_{r}$, is what we see in plot b) of Fig.~\ref{fig:naiive_and_actual_rate}.

\begin{figure}[hbt]
    \begin{minipage}[c]{0.25\textwidth}
    \hspace*{0.8cm}\includegraphics[width=\textwidth]{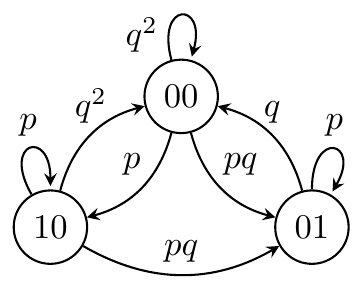}
  \end{minipage}\hfill
  \begin{minipage}[c]{0.65\textwidth}
    \caption{
       Markov Chain generated using the BMCM (Basic Markov Chain Method) for the parameters $n=2$ and $d=1$. The parameter $p$ is the bin occupancy probability, and $q=1-p$. We have three states corresponding to frame configurations 00, 01, and 10 and labeled accordingly. Note how the downtime prevents us from ever being able to observe the 11 frame configuration, and therefore we do not include it in this Markov Chain.
    } \label{fig:bmcm_toy}
  \end{minipage}
\end{figure}

One way to build an IMC is to assign each possible frame configuration to one state. Consider Fig.~\ref{fig:bmcm_toy}.

\begin{figure}[hbt]
    \begin{minipage}[c]{0.6\textwidth}
    \includegraphics[width=\textwidth]{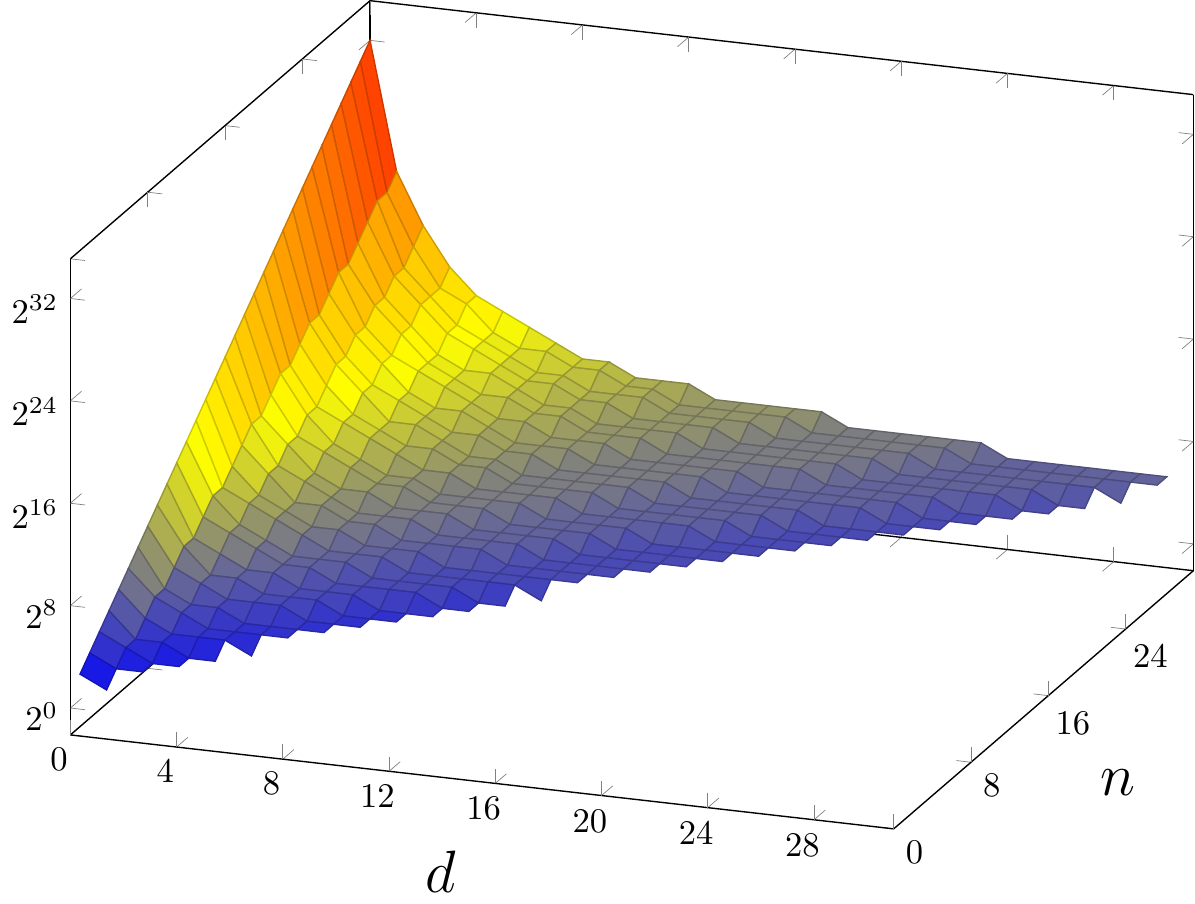}
  \end{minipage}\hfill
  \begin{minipage}[c]{0.3\textwidth}
    \caption{
       The number of states in the Markov Chain generated by the BMCM (Basic Markov Chain Method) as a function of $n$ and $d$. This Markov Chain models the input to the detector on a frame-by-frame basis. Note how the number of states increases rapidly when $n$ increases, especially when $d$ is kept small. Points where $d>n$ are outside the domain of this analysis and so not included in the plot. Points on the plot are colored on a spectrum from blue to red based on their vertical height, which represents the state count.
    } \label{fig:bmcm_states}
  \end{minipage}
\end{figure}

We will call this method the \textit{Basic Markov Chain Method}, or BMCM. If $d=0$ (no downtime) this results in $2^n$ states. With non-zero downtime, some states may be dropped, but the number of states still increases exponentially with $n$. See Fig.~\ref{fig:bmcm_states}.

\label{section:triplet_explanation}

\begin{figure}[hbt]
    \begin{minipage}[c]{0.6\textwidth}
    \includegraphics[width=\textwidth]{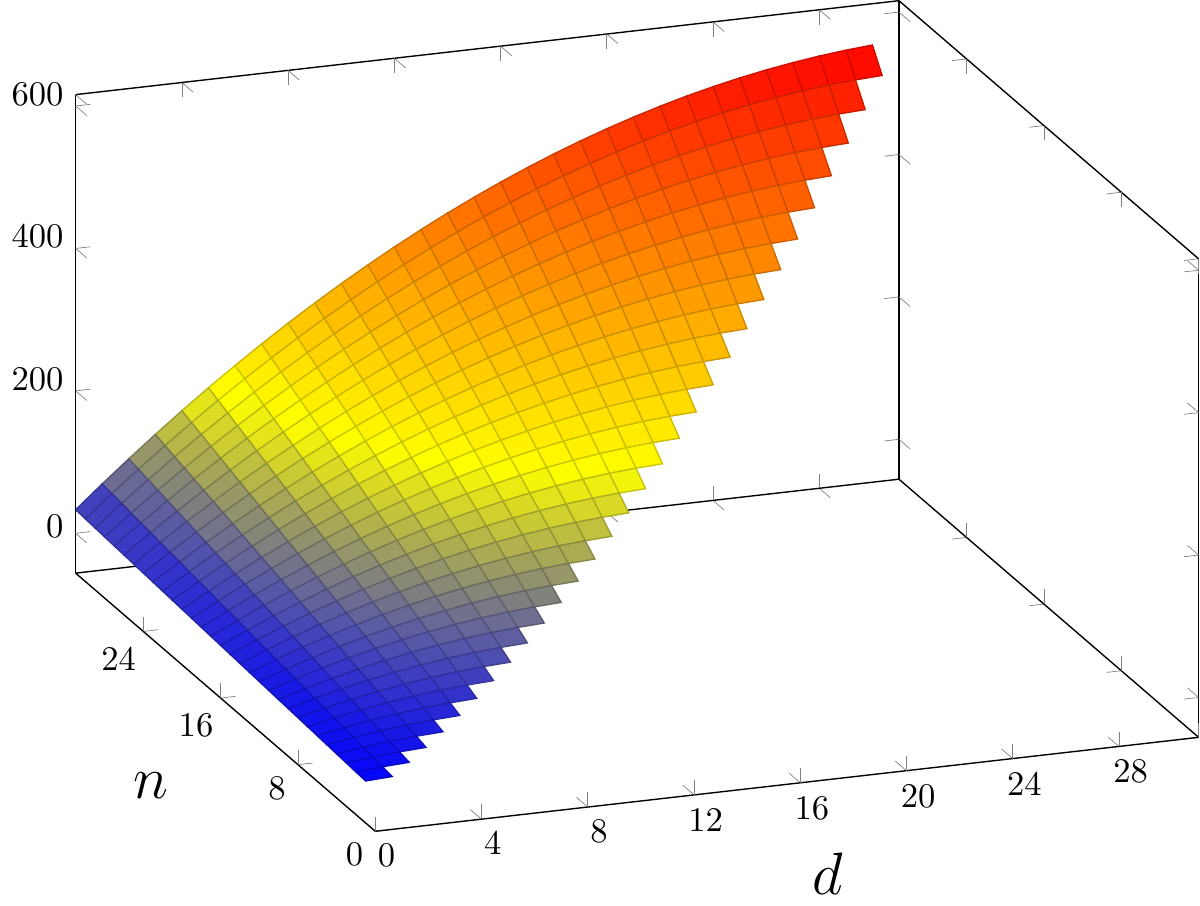}
  \end{minipage}\hfill
  \begin{minipage}[c]{0.3\textwidth}
    \caption{
       The number of states generated in the RMCM (Reduced Markov Chain Method), which is used for modeling the input to the detector on a frame-by-frame basis, as a function of $n$ and $d$. Points where $d>n$ are outside the domain of this analysis and so not included in the plot. Points on the plot are colored on a spectrum from blue to red based on their vertical height, which represents the state count.
    } \label{fig:rmcm_states}
  \end{minipage}
\end{figure}

To avoid this exponential increase in the number of states, we devised an alternative method of assigning frame configurations to Markov Chain states, which we call the \textit{Reduced Markov Chain Method}, or RMCM. In the RMCM, multiple frame configurations can be "combined" and represented by a single, augmented Markov Chain state. These states are augmented in the sense that information about how states are combined is encoded into a triplet of integers $(d_o,n_1,d_i)$ associated with each state, combined or otherwise. We can use these triplets to not only uniquely identify states, but also undo any "compression".
The parameter $d_o$ represents the number of bins that are guaranteed to be non-occupied due to downtime from a previous frame leaking into the current frame. The parameter $d_i$ is the number of bins of downtime leaking out of the current frame into the next frame. With $d_o$ and $d_i$, we can describe the occupancy of bins on the edge of the frame. For instance, consider when $d=5$ and $d_o=3$. It must then be the case that the last two bins are unoccupied, and the third bin from the last must be occupied.
Now we can understand the parameter $n_1$, which is the number of occupancies inside the remaining space of bins, which we haven't yet been fixed by $d_o$ and $d_i$. This is where the compression happens; for instance, if $d=0$ (implying $d_o=0$ and $d_i=0$ as well) and $n_1=1$, then the state associated with the triplet $(0,1,0)$ actually represents $n$ different states at once. This compression results in the number of states increasing polynomially with $n$, instead of exponentially. See Fig.~\ref{fig:rmcm_states}. 

As seen in Fig.~\ref{fig:rmcm_states}, representing states this way has its demerits. Consider the settings $n=2$ and $d=1$. In this case, the Markov Chain generated by the RMCM has five states while the BMCM yields three states. This is because in the RMCM, we actually distinguish between frames where an unoccupied bin is observed through chance and frames where an unoccupied bin is only a result of downtime. For instance, consider the frame configuration "01". It could be that there was no downtime in the front of the frame, in which case we will see this frame configuration with probability $pq$. However, suppose there was some downtime -- perhaps from a previous frame -- that had leaked into this frame. In that case, the first unoccupied bin is guaranteed, so the probability of seeing this frame is now just $p$.

\begin{figure}[H]
    \begin{minipage}[c]{0.45\textwidth}
    \includegraphics[width=\textwidth]{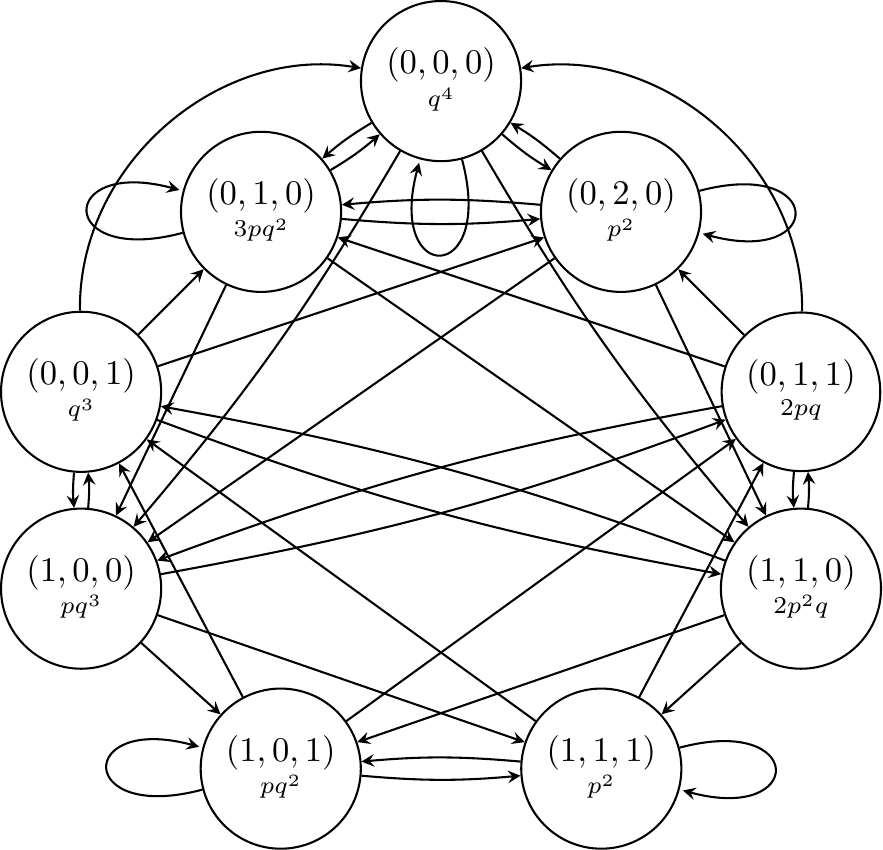}
  \end{minipage}\hfill
  \begin{minipage}[c]{0.45\textwidth}
    \caption{
       Markov Chain generated using the RMCM (Reduced Markov Chain Method) for the parameters $n=4$ and $d=1$. The parameter $p$ is the probability that a bin is occupied, and $q=1-p$. Each state is labeled by a triplet of integers, $(d_o,n_1,d_i)$, as described in section \ref{section:triplet_explanation}. Instead of showing transition probabilities on each edge, the transition probability is shown on the state itself, under its $(d_o,n_1,d_i)$ identifier.
    } \label{fig:rmcm_toy_2}
  \end{minipage}
\end{figure}

When we bump up $n$ to 4 while keeping $d$ at 1, the RMCM starts to close the gap. Consider Fig.~\ref{fig:rmcm_toy_2}; this Markov Chain has nine states, while the Markov Chain generated by the BMCM would have eight states. More importantly, we start to see the first signs of states being "combined".

Consider the state in Fig.~\ref{fig:rmcm_toy_2} labeled $(0,1,0)$. This single state represents the three frame configurations, "1000", "0100", and "0010", where we have no downtime going into the frame. Note how all frames represented by the state have only a single occupied bin and thus will yield $\log_2 4 =2$ bits. This is not a coincidence; the way states can be "compressed" was designed with PPM in mind. In this way, the triplet of numbers each state is labeled with is enough to tell us how many bits, on average, the frames represented by that state will yield. This way, we do not have to go through the trouble of undoing the compression and checking each frame configuration one by one, which would partially defeat the purpose of using the RMCM as an alternative to BMCM for calculating $R_{\text{raw}}$.

\begin{figure}[H]
    \begin{minipage}[c]{0.45\textwidth}
    \includegraphics[width=\textwidth]{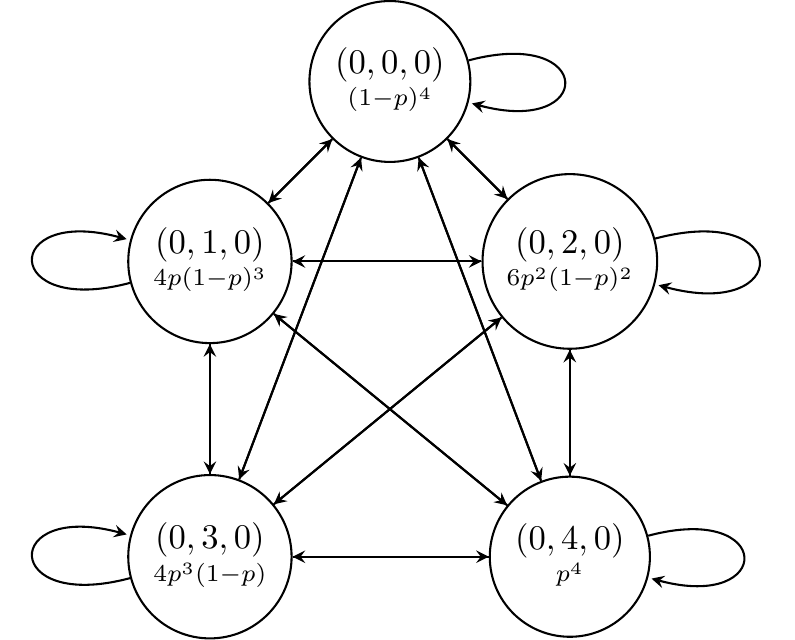}
  \end{minipage}\hfill
  \begin{minipage}[c]{0.5\textwidth}
    \caption{
       Markov Chain generated using the RMCM (Reduced Markov Chain Method) for the parameters $n=4$ and $d=0$. The parameter $p$ is the bin occupancy probability and $q=1-p$. Each state is labeled by a triplet of integers, $(d_o,n_1,d_i)$, as described in section \ref{section:triplet_explanation}. Instead of showing transition probabilities on each edge, the transition probability is shown on the state itself, under its $(d_o,n_1,d_i)$ identifier. The BMCM would have generated 16 states, but here we have 5 states instead.
    } \label{fig:rmcm_compress_good}
  \end{minipage}
\end{figure}

For an example where the RMCM produces fewer states than the BMCM, see Fig.~\ref{fig:rmcm_compress_good}.

\begin{figure}[hbt]
    \begin{minipage}[c]{0.6\textwidth}
    \includegraphics[width=\textwidth]{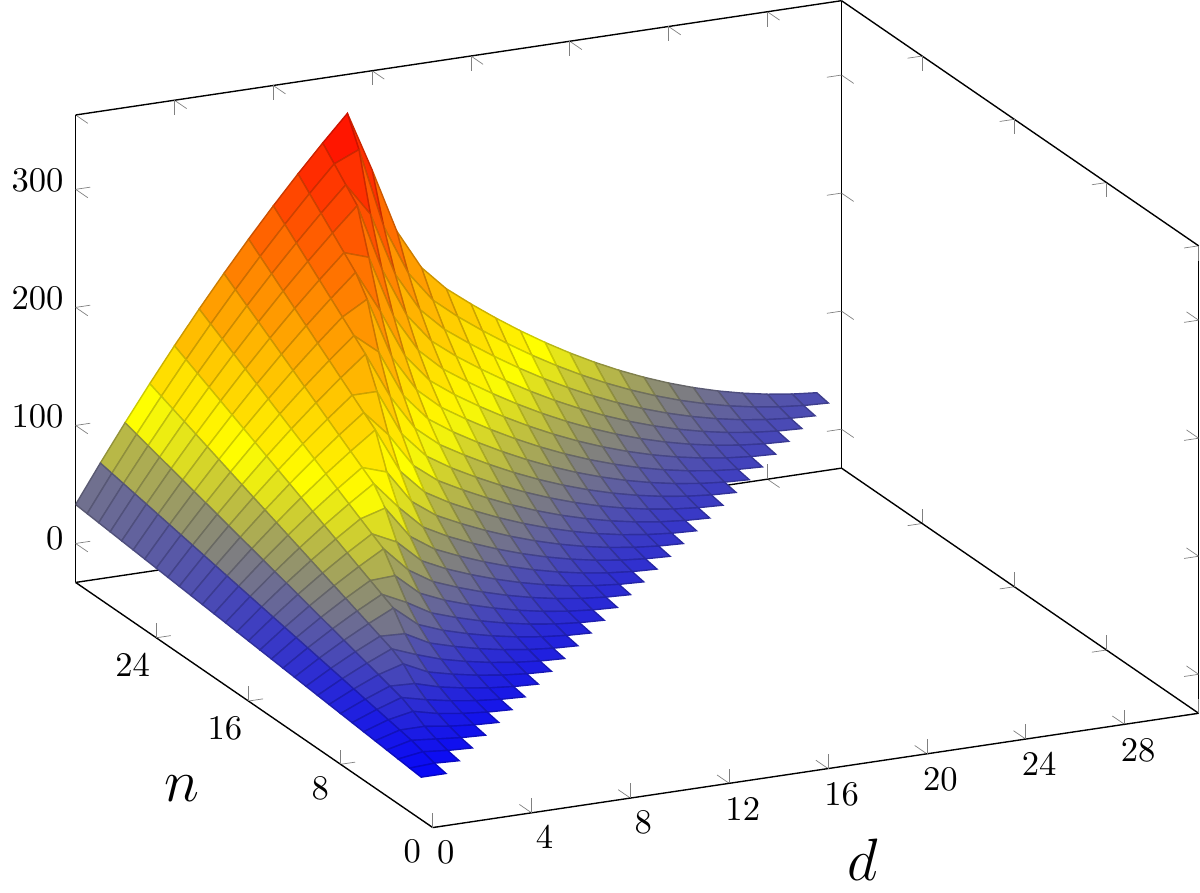}
  \end{minipage}\hfill
  \begin{minipage}[c]{0.3\textwidth}
    \caption{
       A 3D plot of the minimum number of states generated in the BMCM (Basic Markov Chain Method) versus the RMCM (Reduced Markov Chain Method) as a function of $n$ and $d$. Note how the two techniques appear to complement each other, producing a smaller number of states where the other produces a large amount. A ridge exists appearing to lie on the line where $d=n/2$, representing the boundary where the BMCM and RMCM generate approximately the same number of states. Points where $d>n$ are outside the domain of this analysis and so not included in the plot. Points on the plot are colored on a spectrum from blue to red based on their vertical height, which represents the state count.
    } \label{fig:bmcm_vs_rmcm_states}
  \end{minipage}
\end{figure}

Notice how in the example where the RMCM does better than the BMCM, we \textit{decreased} the parameter $d$ from 1 to 0. After doing so, the number of states generated by the BMCM decreases as $d$ increases, but the number of states generated by the RMCM increases. Therefore, we consider both methods in a hybrid method wherein we select the method that results in the fewest number of Markov Chain states. It seems, based on inspection of the data used to produce Fig.~\ref{fig:bmcm_vs_rmcm_states}, that a good enough rule is to choose BMCM when $d>n/2$ and choose RMCM otherwise. This may not always be the case though, so if we absolutely must choose the more efficient method, one may utilize Eq.~\ref{eq:bmcm_states} and Eq.~\ref{eq:rmcm_states}, which give the exact number of states for the BMCM and RMCM, respectively.

The number of states used in the BMCM when given $n$ and $d$ is given by the recursive function $N_{\text{BMCM}}$ as defined in Eq.~\ref{eq:bmcm_states}.
\begin{equation}
    \label{eq:bmcm_states}
    N_{\text{BMCM}}(n,d) = \begin{cases}
        1 & n<1 \\
        N_{\text{BMCM}}(n-1,d)+N_{\text{BMCM}}(n-1-d,d) & n\geq 1
    \end{cases}.
\end{equation}

The number of states used in the RMCM when given $n$ and $d$ is described by the function $N_{\text{RMCM}}$ as defined in Eq.~\ref{eq:rmcm_states}.

\begin{equation}
    N_{\text{RMCM}}(n,d) = \sum_{d_i=0}^d \left(N'(d_i,0)+
    \sum_{d_o=\max(1+d+d_i-n,1)}^d N'(d_i,d_o)\right)
    \label{eq:rmcm_states}
\end{equation}
where
\[
    N'(d_i,d_o) = \bigg\lfloor\frac{ n'(d_i,d_o)}{d+1}\bigg\rfloor+1
~ \text{and} ~
    n'(d_i,d_o) = \begin{cases}
        n-d_i & d_o=0 \\
        (n-d_i)-(d+1-d_o) & d_o>0
    \end{cases}.
\]

With an IMC in hand, we can calculate $R_{\text{raw}}$. Suppose we know the stationary probability of each state. In that case, we also know how likely the detector is to observe a certain frame configuration (since each state is mapped to one or more frame configurations). We will also know how many bits that frame configuration will get us if we know what binning scheme we are using. We can thus calculate a weighted average to get $R_{\text{raw}}$. In this context, the stationary probability is the weights, and bits from each frame configuration are averaged.

\begin{figure}[hbt]
    \begin{minipage}[c]{0.3\textwidth}
    \includegraphics[width=\textwidth]{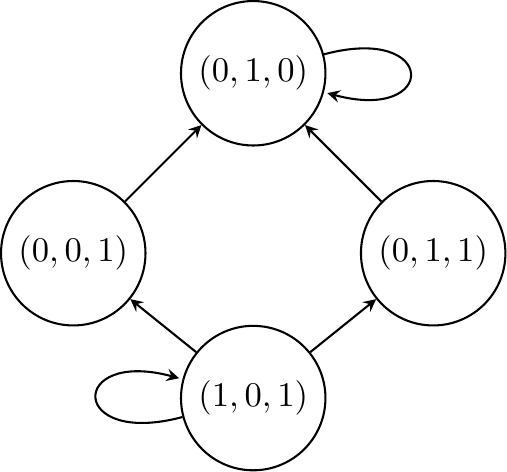}
\end{minipage}\hfill
\begin{minipage}[c]{0.65\textwidth}
    \caption{
       Output Markov Chain (OMC) for the parameters $n=4$ and $d=1$, for the PPM binning scheme. Each state is labeled by a triplet of integers, $(d_o,n_1,d_i)$, as described in section \ref{section:triplet_explanation}. Transition probabilities are not shown. The method we use to derive the transitions begins with a state vector $\pi$ for the IMC, where $\pi_i=1$ when $i$ corresponds to an IMC state that yields a non-zero number of bits. We apply the original transition matrix once. Then we switch to another transition matrix where all states that give a non-zero number of bits transition to themselves with probability 1. We then find the stationary probability using this alternate transition matrix. At the end of this step, all state vector elements $\pi$ should be zero if the element corresponds to a state that yields no bits. The resulting probabilities of each state that yields bits are the transition probabilities from the state in the OMC corresponding to state $i$ in the IMC to the corresponding states in the OMC.
    } \label{fig:omc_toy}
  \end{minipage}
\end{figure}

However, to ensure the uniform randomness of the key, we need to find $C^d_{r}$, which will tell us by how much we need to compress the resulting bitstream to achieve perfect randomness. We now need to generate an OMC from the IMC. This involves mapping each Markov Chain state to a new state -- or states -- that correspond to any output bits. We will also need to find the stationary entropy of this new Markov Chain since we need it for the Markov Chain Entropy formula. This jump from the IMC to the OMC is necessary because if we applied the Markov Chain Entropy formula to an IMC, we would recover Eq.~\ref{eq:basic_upper_bound}. To see how this may be done, consider Fig.~\ref{fig:rmcm_toy_2} (an IMC) and Fig.~\ref{fig:omc_toy} (its OMC).

For more information on the derivation of an OMC, we direct the reader to \cite{Cheng:Thesis:2022}.

Recall that the RMCM is useful if we are using PPM because we can tell how many bits every frame configuration represented by a state yields without "decompressing" it. The situation is similar to the OMC, where things are fine as long as we use PPM because, with PPM, every single frame in a state gets the same amount of bits (so instead of average, we really should say that the number we get is \textit{the} number of bits we get for that state). More importantly, a unique input frame configuration gives that output for every possible binning scheme output, which is not the case for Simple Binning. For example, both 1000 and 0111 would yield the same output bits. These properties of PPM give us the luxury of not having to decompress the IMC to derive an accurate OMC. However, in the general case, more sophisticated techniques are required when we consider newer, more advanced binning schemes such as those described in \cite{QKD:KarimiSW20}. In the worst case, the advantage of using the RMCM is not only negated but even subverted, since we will have to decompress it straight away in order to calculate even $R_{\text{raw}}$, but the "base" Markov chain is larger than that of the BMCM.

\subsection{Observations}

\begin{figure}[hbt]
    \begin{minipage}[c]{0.4\textwidth}
    \includegraphics[width=\textwidth]{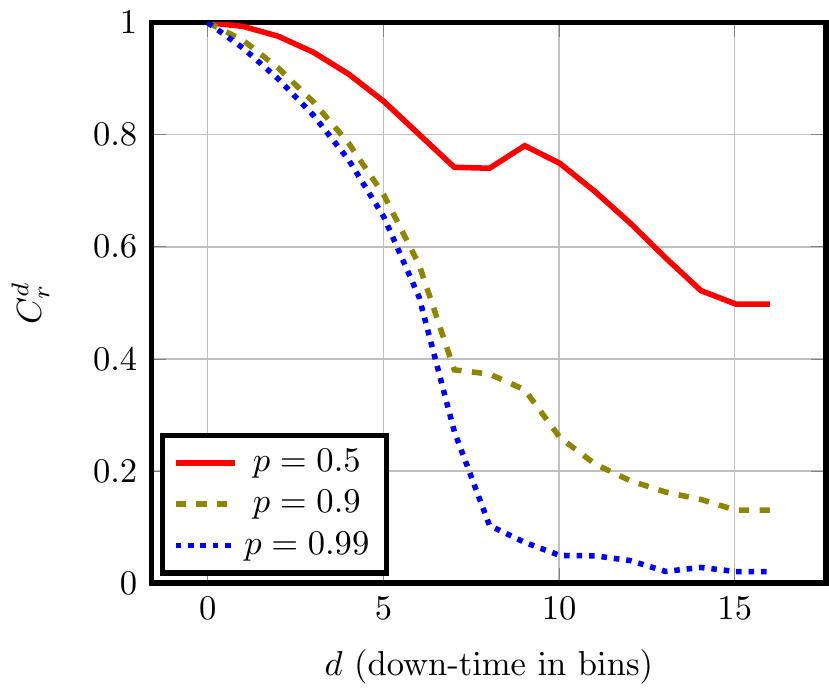}
  \end{minipage}\hfill
  \begin{minipage}[c]{0.5\textwidth}
    \caption{
       $C^d_{r}$ for $n=16$ and when $d$ varies over the x-axis. Values for $p$ are chosen by how well they show the change in the curve versus the change in the value of $p$ (word this better). This shows how $C^d_{r}$ is increasingly affected by non-zero downtime as $p$ approaches 1. The data was generated using 
    our online tool that can be found at \url{https://cc1539.github.io/qkd-binning-demo-2/}
    } \label{fig:dn_ratio}
  \end{minipage}
\end{figure}

Recall how when $p$ is small, $C^d_{r}$ remains close to 1. This makes sense because in that case, photon inter-arrival times are, on average, spaced apart enough for downtime not to have a chance to change anything. It is when $p$ approaches 1 (when $\lambda_p$ is increased arbitrarily) that $C^d_{r}$ falls. See Fig.~\ref{fig:dn_ratio}.

For an intuitive explanation, consider a toy example where $n=2$, $d=1$, and $p=1$. In this case, the detector will observe the sequence 1010101010 repeated indefinitely. Notice how since $n=2$, every frame would be a valid one-shot encoding, which gives us one bit per every two bins, and if we did not know better, we would think this is great. However, the frame itself is repeating, so we repeatedly get that same bit, resulting in a perfectly deterministic sequence of all 0s or 1s. As long as $d$ is non-zero, the same effect applies for all $p$, but usually much less depending on $p$.

\begin{figure}[hbt]
    \begin{minipage}[c]{0.45\textwidth}
    \includegraphics[width=\textwidth]{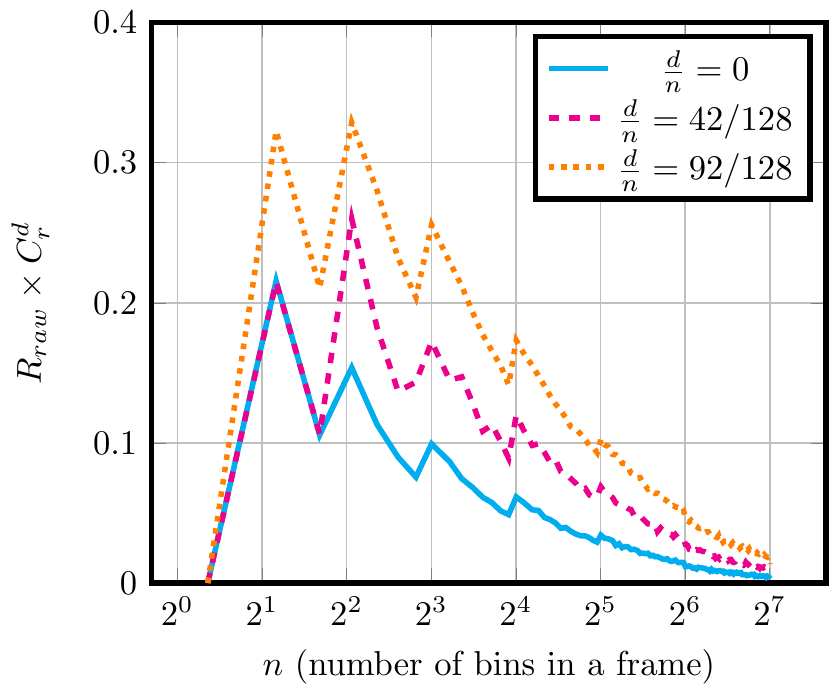}
  \end{minipage}\hfill
  \begin{minipage}[c]{0.45\textwidth}
    \caption{
       $R_{\text{raw}}\times C^d_{r}$ for $\lambda_p\approx\frac{2.3}{\tau_b}$, with $n$ varying over the x-axis from 0 to 128. Here, frame width in time is kept constant, such that if $T_f$ is the duration of a frame, then $\tau_b=T_f / n$. The parameter $d$ is also kept constant with respect to time, so the ratio $\frac{d}{n}$ is also constant. This expression for $\lambda_p$ corresponds to a value of $p=0.9$ for the case of $n=1$. We are using PPM. The data was generated using our online tool available at the following URL: \url{https://cc1539.github.io/qkd-binning-demo-2/}
    } \label{fig:n_vary}
  \end{minipage}
\end{figure}

If downtime was zero, we'd never see any of these decreases in $C^d_{r}$ no matter how high we push $\lambda_p$. Does this mean we want $\frac{d}{n}$ to be as small as possible? This may not the case. This time, while considering $C^d_{r}$ we instead neglected $R_{\text{raw}}$. See Fig.~\ref{fig:n_vary}.


Notice how $R_{\text{raw}}\times C^d_{r}$ seems to actually be increasing as $\frac{d}{n}$ increases. To be clear, we \textit{are} taking into account the loss of entropy due to downtime here, and yet the ultimate key rate is still going up as $d$ increases. This indicates that somehow, the increase in the number of key bits caused by increasing downtime wins out against the decrease in the number of bits lost due to the decrease in $C^d_{r}$. So does this mean that we want $d$ to be as large as possible? Perhaps we may want to change the binning scheme instead. Today's specifications (namely, $\lambda_p$ being quite low) mean that we are not affected by downtime too seriously -- recall that if $p$ is small enough, $d$ stops playing much of a role. However, as we can achieve greater $\lambda_p$, we may need newer, more efficient bit extraction techniques. Such schemes, which unlike PPM do not discard all frames with multiple arrivals, have been proposed \cite{QKD:KarimiSW20}.

\section{Modeling the Impact of Dark Counts} 
\label{sec:darkcounts}

Occasionally, an SPD announces a photon arrival independent of an SPDC-generated photon event. This phenomenon is known as a dark count. Primarily, dark counts are due to light leakage into the quantum channel. However, even the best systems report a non-negligible dark count rate. Dark counts can cause two types of errors. In Type-One errors a dark count occurs within the same time frame as an SPDC-generated pair and invalidates an otherwise acceptable PPM frame. Thus, the frame's contribution to the key is lost. In Type-Two errors, frame-coincident dark count detections happen at the two stations, which causes Alice and Bob to retain frames with uncorrelated arrivals that are not due to entangled photons but are indistinguishable to them from such frames.

This section aims to model the impact of dark count arrivals on the system in light of these two potential types of errors. We first show the natural result that dark counts can decrease the probability of observing a PPM-valid time frame. We then model the frame-coincident dark count arrivals as large jitters. Type-One errors reduce the key rate by decreasing the number of PPM-valid frames. Type-Two errors reduce the key rate by forcing Alice and Bob to communicate more during the privacy amplification stage to reconcile the differences even in their uncorrelated frames.

We model the dark count arrival process at Alice/Bob's measurement station as a Poisson process with rate parameter $\lambda_{\text{DC}}$. Recall that the arrival process of SPDC-generated photons is also Poisson with rate $\lambda_{\text{p}}$. Therefore, we can determine the probability that Alice and Bob observe a frame to be PPM valid as the total probability that they each observe a frame with a single photon due to either photon source. From the properties of Poisson processes, we know that the probability of a single event, and zero events, within a time frame will respectively follow the form of \begin{equation}
    p_{\text{1}} = \lambda T_f \exp \left(-\lambda T_f\right), \text{and } p_{\text{0}} = \exp \left(-\lambda T_f\right).
    \label{eq:frameoccprob}
\end{equation}

By substituting the respective rate parameters into these two formulas, we can arrive at the total probability of observing the PPM frame described by
\begin{equation}
    p_{\text{PPM}} = 
    p_{\text{SPDC},1}\cdot p_{\text{DC},0}^2 +
    p_{\text{SPDC,0}}\cdot p_{\text{DC},1}^2.
    \label{eq:PPMvalid}
\end{equation}

The first term in the Eq.~\ref{eq:PPMvalid} is the probability that a frame with a single SPDC-generated photon arrival is retained in the absence of Type-One dark count errors. The second term is the probability that a frame is considered valid due only to coincident dark counts. This represents the probability of a Type-Two dark count error. In each of these terms, the dark count probability component is squared to account for the fact that Alice and Bob must both register a dark count to retain the frame as a PPM frame. The SPDC probability is not squared as a single SPDC event generates both photons.
Note that if only Alice's or only Bob's frame is corrupted, they both throw the corresponding frame away. Thus, in any scenario other than those described by the two terms in Eq.~\ref{eq:PPMvalid}, Alice and Bob will throw away the frame.
As the ratio of dark counts to SPDC photons increases, the rate of observing PPM valid time frames drops quickly. The decrease in key rate due to type-one dark count errors follows this same trend. Every frame lost corresponds to the loss of $n$ raw key bits. Due to this phenomenon, it is incredibly important to minimize dark counts when using time-entanglement based QKD with PPM. 

\subsection{Modeling Dark Counts as Jitters} 

Using the probabilities found above, we can describe the \textit{observed} timing jitter PDF as a weighted combination of jitter due to detector timing jitter \textit{and} perceived jitter due to the arrivals of coincident dark counts. With this combined jitter error PDF, we can compute the loss of key rate much like in Section~\ref{sec:inforec}. Here we substitute the double Gaussian PDF with this new symmetric jitter PDF. In doing so, we can include the key loss due to over-correcting for coincident dark counts in the term $\beta_r$ from Eq.~\ref{eq:metrics}. Despite these frames being entirely uncorrelated, if Alice and Bob cannot discern them from SPDC frames then in attempting to correct the errors they leak information about the keys. This reduces the secrecy of the keys including the truly correlated contributions, and thus reduces the reconciled key length. 

The dark counts are uniformly distributed within a given time frame. Thus, to determine the distribution of the observed jitter due to coincident dark counts, we convolve two uniform PDFs. This results in a triangular distribution with a peak at the jitter value 0 described by $\Tri(x) = \frac{1}{T_f} - \frac{1}{T_f^2}\lvert x \rvert$. We consider this as a different type of observed jitter error. We can write the distribution of observed jitter errors as a weighted sum of the distribution of detector jitter errors and jitter errors due to coincident dark counts. The weighting term is a function of the frame size as it represents the ratio between the probability that a frame is PPM valid due to an SPDC photon pair to the total probability that a frame is PPM valid. Given a specific frame size parameter and the expected photon arrival rates, we get
\begin{align}
        p(t_j) &= c(T_f) N(0, 2\sigma_d^2) + (1-c(T_f)) \Tri(t_j)
    \label{eq:pjitter}\\
        c(T_f) & = \frac{p_{\text{SPDC,1}}\cdot p_{\text{DC,0}}^2}{p_{\text{PPM}}}
        = \frac{p_{\text{SPDC,1}}\cdot p_{\text{DC,0}}^2}{ p_{\text{SPDC},1}\cdot p_{\text{DC},0}^2 +
p_{\text{SPDC,0}}\cdot p_{\text{DC},1}^2}.
\end{align}

\begin{figure}[hbt]
    \begin{minipage}[c]{0.4\textwidth}
    \includegraphics[width=\textwidth]{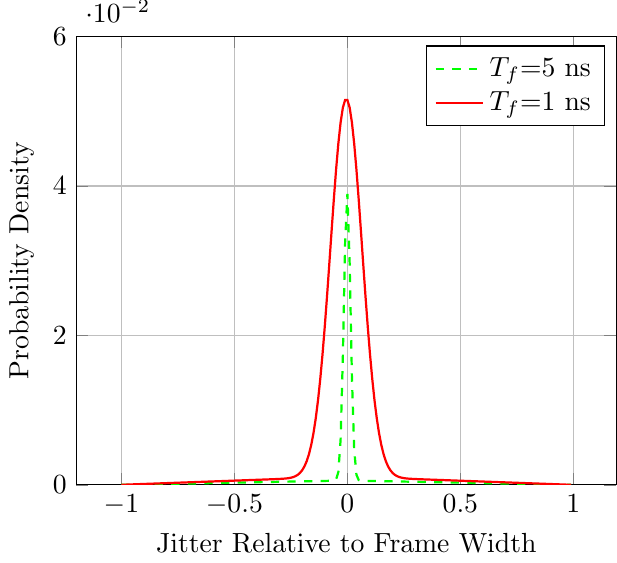}
  \end{minipage}\hfill
  \begin{minipage}[c]{0.5\textwidth}
    \caption{Two frame configurations showing the theoretical model of observed jitter errors. A frame size of 5 ns was chosen for the dashed plot and 1 ns for the solid plot to illustrate the relative differences. These plots were generated using Eq.~\ref{eq:pjitter}.} \label{fig:channelCharacterization}
  \end{minipage}
\end{figure}

We can see the effect of considering coincident dark counts on the jitter PDF in Fig.~\ref{fig:channelCharacterization}. Without considering dark counts, there is some motivation to increase time frame width to allow for increasing the width of time bins while maintaining a high bin count per frame. This approach would intuitively reduce the number of errors due to detector jitter, as any jitter would be more likely to land within the same bin. Here we see that if you are not careful in doing so, you can face different jitter errors due to an increased susceptibility to coincident dark counts. The Gaussian seems to dominate with smaller frame sizes, but we can observe heavier tails to the distribution with larger time frames. These heavy tails indicate a higher frequency of significant jitter errors resulting from the coincident dark count frames.

\begin{figure}[hbt]
    \begin{minipage}[c]{0.4\textwidth}
    \includegraphics[width=\textwidth]{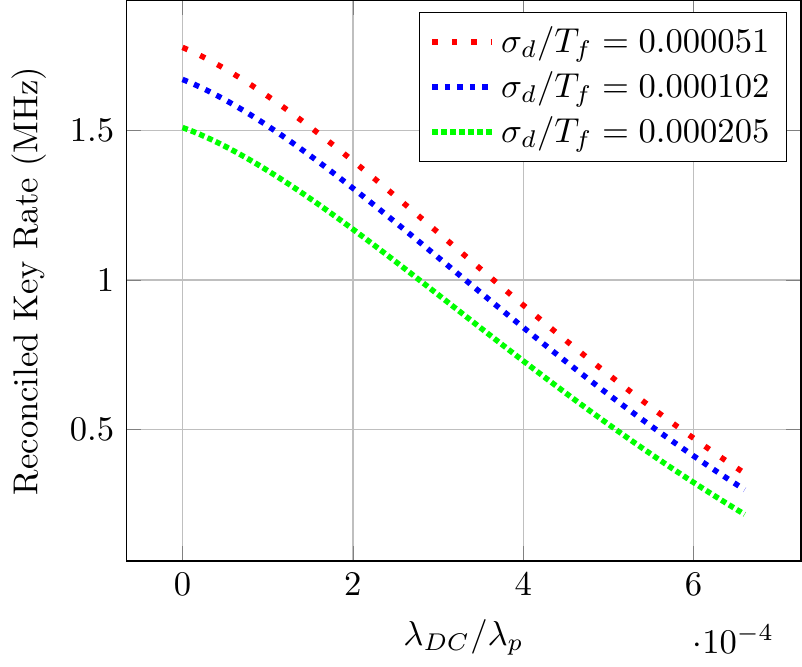}
  \end{minipage}\hfill
  \begin{minipage}[c]{0.5\textwidth}
    \caption{ These plots were generated using equations \ref{eq:frameoccprob}-\ref{eq:pjitter} to replace equation \ref{eq:tp} with the same methods as used in Fig.~\ref{fig:postreconcilliationrates}. Here the reconciled key rate is labeled in MHz as we account for the probability that a frame is considered PPM valid. This is the key rate in time versus the earlier discussed key rate per PPM valid time frame.}
    \label{fig:perrdc4}
  \end{minipage}
\end{figure}

Finally, we applied the methods in Sec.~\ref{sec:inforec} with the above new jitter model to calculate the secret key rate loss due to information reconciliation. Fig.~\ref{fig:perrdc4} shows the resulting plot in which we can see the post-reconciliation key rate in MHz as a function of the relative dark count rate to the SPDC generation rate. This effect is compounded by the reduced frame utilization, which reduces the post-reconciled key rate further. This evident trade-off further motivates the need for increasing the SPDC generation rate. However, if increased too much, the downtime effects may come into play. In the short term, these findings may be used to quantify expected experimental key rates given practical implementation details. 

\section{Conclusion\label{sec:conclusions}}

Time-entanglement-based QKD seems a natural alternative to standard QKD when the goal is to increase the secret key rate in limited photon generation rates. However, in this model, some potentially catastrophic imperfections threaten to reduce the secrecy of generated keys. In this paper, we characterized several pitfalls related to time-of-arrival QKD. We provided mechanisms to accurately predict the possible secret key rate given the degree an experiment suffers from each impairment. First, we considered how timing measurement jitter introduces errors and necessitates information reconciliation, which decreases key length. Second, we accounted for the loss of entropy in generated keys due to detector downtime, which introduces memory to the system. We also analyzed dark counts, assuming that Alice and Bob cannot distinguish coincident dark count frames from SPDC photon arrival frames. 

One of our most important contributions was developing a novel method to calculate entropy loss due to downtime. Under specific parameters, it is possible to simplify the Markov chain model of photon arrivals at a detector given detector downtime. This reduced model is more computationally tractable, allowing us to compute entropy reduction when working with frames with many time bins. An interactive online tool was developed in conjunction with this approach to enable researchers to explore the effect of binning scheme parametrization on the secret key rate in the light of downtime (\href{https://cc1539.github.io/qkd-binning-demo-2/}{https://cc1539.github.io/qkd-binning-demo-2/}).

Future research in this space can expand on many of these aspects. Instead of considering various impairments separately and combining them afterward, one could think of them simultaneously. We see another interesting angle to consider joint information reconciliation and privacy amplification schemes. We hypothesize that compressing the raw key once for information reconciliation and again for privacy amplification may be overcompensating in some scenarios. That is, the value we would get from \ref{eq:final} -- using parameters $C^d_r$ and $\beta_r$ as calculated in the ways we've described so far -- may be lower than it needs to be. To reframe: we study information reconciliation in the context of jitter and privacy amplification in the context of downtime. We hope that accounting for jitter in our downtime analysis (which currently assumes a noiseless channel) may allow us to pull $C^d_r$ closer to 1 without sacrificing key entropy.
Additionally, we see potential in more advanced error-correcting codes for the information reconciliation phase. Specifically, we expect that layered codes will do well at dealing with multiple sources of errors. Jitter errors may be corrected separately from the coincident dark count errors, reducing the cost of information reconciliation on the secret key rate.

\appendices

\section*{Acknowledgment}
We thank the following colleagues:
Murat Can Sar{\i}han for providing and explaining experimental data, Esmaeil Karimi and Phil Whiting for general discussions on non-ideal detectors and Pei Peng for his feedback on an earlier version of this paper.

\ifCLASSOPTIONcaptionsoff
  \newpage
\fi

\bibliographystyle{ieeetr}
\bibliography{bibio}

\end{document}